\documentclass[preprint,aps,amsmath,amssymb,superscriptaddress]{revtex4}

\usepackage[cp1250]{inputenc}

\newcommand{\ket}[1]{\mid \! #1 \rangle}

\begin{document}

\title{Quantum volume and length fluctuations in a midi-superspace model of Minkowski space}

\author{Jeremy Adelman}
\affiliation{Department of Physics, University of California,
One Shields Avenue, Davis CA 95616 USA}
\author{Franz Hinterleitner}
\affiliation{Department of Theoretical Physics and Astrophysics,
Faculty of Science of the Masaryk University, Kotl\'{a}\v{r}sk\'{a}
2, 611\,37 Brno, Czech Republic}
\author{Seth  Major}
\affiliation{Department of Physics, Hamilton College, Clinton NY
13323 USA}

\date{January 2015}

\begin{abstract} In a (1+1)-dimensional midi-superspace model for
gravitational plane waves, a flat space-time condition is imposed
with constraints derived from null Killing vectors. Solutions to a
straightforward regularization of these constraints have diverging
length and volume expectation values. Physically acceptable
solutions in the kinematic Hilbert space are obtained from the
original constraint by multiplying with a power of the volume
operator and by a similar modification of the Hamiltonian
constraint, which is used in a regularization of the constraints.
The solutions of the modified Killing constraint have finite expectation
values of geometric quantities. Further, the expectation value of
the original Killing constraint vanishes, but its moment is non-vanishing.
As the power of the volume grows the moment of the original
constraint grows, while the moments of volume and length both
decrease. Thus, these states provide possible kinematic states for flat space, with
fluctuations. As a consequence of the regularization of
operators the quantum uncertainty relations between geometric
quantities such as length and its conjugate momentum do not reflect
naive expectations from the classical Poisson bracket relations.\\

\end{abstract}
\maketitle

\section{Introduction}

Loop quantum gravity (LQG) quantizes the spatial geometry by
introducing ``atoms of spatial geometry" in form of quanta of volume, area,
length, and angle \cite{RSareavol,Lvol,ALareavol,Tlength,Mangle,Blength}. Unlike the Minkowski vacuum in quantum field
theories of different kinds of matter, the quantum
model of flat space appears to be, not a ``no particle" state, but rather a highly excited
state with a macroscopically homogeneous distribution of excited quanta of
geometry. In this paper we explore the nature of quantum flat space in an effectively (1+1)-dimensional gravitational
system tailored to study the propagation of plane gravitational waves. This is the third paper in a series \cite{HM,HM2}
on the quantization of gravitational
plane waves with the eventual goal of quantizing ``small amplitude" plane gravitational waves
on flat space and to ascertain the effects (if any) of the underlying fundamental geometric discreteness of LQG on the propagation of waves.
This work, identifying candidate kinematic states of flat space, is a step toward that goal.

One advantage of the present approach is that we can derive model
states from classical flatness conditions, in the form of constraints on
quantum states derived from the existence of Killing vector fields.
Additionally, the resulting algebra of constraints, including these new ``Killing constraints", 
is first class \cite{HM2}. The second, main advantage of the present -- so far
kinematical -- model is the realistic chance to subject them to
quantum dynamics. For, even if the midi-superspace Hamiltonian constraint is not simple,
it is not so complicated that, from the very beginning, it thwarts
the application of the dynamics to the candidate flat space states described here.

One important aspect of the problem that this paper does not address is the derivation of
physical states.  Although the present quantization uses the Hamiltonian constraint in 
formulating quantum kinematic constraints, we do not obtain physical solutions or check that the constraint 
algebra is anomaly-free.  Work on this  is underway.
 
The plane wave class of pp-wave space-times considered here
are derived from the cosmological Gowdy model, which
was quantized by Banerjee and Date \cite{BD} using formal tools from
earlier work by Bojowald and Swiderski \cite{2}. Our earlier work on plane gravitational waves
\cite{HM2}, shows that left- or right-moving wave space-times can be
found using a system of first-class constraints. In this approach a description of
background flat space for wave propagation arises in a natural way.

Earlier work  addressed similar models of gravitational waves.  Neville considered the quantization of plane gravitational waves both within geometrodynamics and with complex connections \cite{N}. Borissov  studied plane waves and weave states \cite{Bweaves}. To quantize a similar model Beetle exploited the observation that the symmetry reduction of non-compact toroidally symmetric space-times yields a system equivalent to a free massless scalar field on a fixed $(2+1)$-dimensional background \cite{B}. Using metric variables Mena Marug\'an and Montejo reduced the model at the classical level using gauge choices and symmetry reduction \cite{MM}. These quantizations leave the relation between the fundamental discrete geometry of LQG and classical local Lorentz invariance veiled. Since this relation is precisely what we wish to elucidate, we take an approach to quantization closer to that of $(3+1)$-dimensional  LQG.

The organization of the paper is the following: In the next section and
Section \ref{QSO} we briefly present
 those basics of the quantized Gowdy model that are
necessary for our adaptation to gravitational waves --
opening up the global toroidal topology to
flat space and finding a set of first-class ``Killing constraints"
that select unidirectional gravitational waves \cite{HM,HM2}.
The geometric quantities used in the analysis of the
flat space constraints are defined in Section \ref{GQ}.
In this section we also interpret the constraints geometrically in terms of
the rate of change of cross section areas and in terms of length.
We show in \ref{geom_ops}
that a set of simple ``vanishing curvature" constraints yields
non-normalizable states in the kinematic Hilbert space, further
motivating the use of the constraints derived from the Killing vectors.

We select quantum states for flat geometry by imposing a
``no wave" constraint derived from the left- and right-moving constraints of Ref. \cite{HM2}.
These Killing constraints, formulated and implemented in Sections \ref{KC} and \ref{kill_c_ops},
suppress all waves and thus give ``no-wave states", a kinematic model for flat space.
Calculating expectation values and fluctuations in geometric quantities around these solutions,
we find that requiring finite expectation values of geometric quantities limits
the formulation of the ``no wave" constraint. 

The main work of this paper is dedicated to
the construction and analysis of the candidate flat space states, 
as detailed in Sections \ref{kill_c_ops} and \ref{HKC}, as well as the appendices.
The constraint operators contain explicit connection components,
which cannot be directly promoted to quantum operators, and must be
regularized. We present two different strategies to do
this, one by applying ``Thiemann's trick" \cite{5} to replace the
constraint by the commutator of a part of the Hamiltonian constraint
with the volume operator, and the other one by directly approximating the
connection in terms of corresponding holonomies and obtaining a Hermitian
constraint. In both cases it turns out that the ``no-wave constraints" in their
straightforward form are too strong -- they produce states with
diverging length and volume expectation values, {\em even} for a single spin network node or vertex.
We call this smallest possible unit an ``atom of geometry". 

We consider two different ways of relaxing the no-wave constraints using the volume operator.
These have normalizable solutions with finite expectation values
and uncertainties for geometric quantities on a single atom of geometry,
giving candidate states for flat space as subset of the kinematical state space of a single
atom of geometry. The kinematical Hilbert space of the whole system is subtile due
to the open topology. (See the work by Thiemann and Winkler \cite{OT}).  We have not 
addressed the normalizability of the constraints in that setting.

\section{Variables and Constraints}
\label{variables}

The space-time model for plane gravitational waves
propagating in the $z$-direction, where the $x$ and the $y$
direction form a plane of homogeneity, is formally very close to the
polarized Gowdy model in Ref. \cite{BD}. The difference lies only in
the global topology; locally both models are formulated in the same
Ashtekar-type variables. Following Ref. \cite{BD}, we introduce
densitized triads in a space-like hypersurface with the component
${\cal E}(z)$ in the inhomogeneous $z$-direction and the homogeneous
transverse components arranged as two-vectors
\begin{equation}\vec{E}^x=(E^x\cos\eta,E^x\sin\eta), \hspace{5mm}
\vec{E}^y=(-E^y\sin\eta,E^y\cos\eta).
\end{equation}
We consider only polarized gravitational waves, where these two
vectors are orthogonal. Like $\cal E$, the components $E^x$, $E^y$,
and $\eta$ are functions of $z$. In terms of these variables the
spatial metric is given by
\begin{equation}\label{sm}
{\rm d}s^2={\cal E}\frac{E^y}{E^x}\,{\rm d}x^2+{\cal
E}\frac{E^x}{E^y}\,{\rm d}y^2+\frac{E^xE^y}{\cal E}\,{\rm d}z^2.
\end{equation}

The canonically conjugate variable to $\cal E$ is the
Ashtekar-Barbero connection ${\cal A}(z)$, conjugate variables to
$E^x$ and $E^y$ are $X(z)$, and $Y(z)$, the extrinsic curvature
components $K_x$ and $K_y$, rescaled by multiplication with the
Barbero-Immirzi parameter $\gamma$. The angular variable $\eta$
represents a pure gauge degree of freedom. Its conjugate momentum
$P_\eta(z)$ is the generator of $U(1)$ rotations in the $(x,y)$
plane. The non-vanishing Poisson brackets are
\begin{eqnarray}
&&\{{\cal A}(z),{\cal E}(z')\}= \{\eta(z),P_\eta(z')\}=
\{X(z),E^x(z')\}=\{Y(z),E^y(z')\}\nonumber\\
&&=\kappa\gamma\delta(z-z')
\end{eqnarray}
where $\kappa$ is proportional to Newton's constant \footnote{In
Ref. \cite{HM} we used $\kappa' = 4 \pi G/A_o$ for the fiducial area
in the transverse plane, but the form of this constant will play no
role in the present work.}. Given the symmetries
of the model the standard gauge-generating constraints reduce to the Gau\ss\ constraint
\begin{equation}
G=\frac{1}{\kappa\gamma}\,({\cal E}'+P_\eta),
\end{equation}
the diffeomorphism constraint
\begin{equation}
C=\frac{1}{\kappa\gamma}\left[X'E^x+Y'E^y-{\cal E}'{\cal
A}+\eta'P_\eta\right],
\end{equation}
and the Hamiltonian constraint,
\begin{equation}\label{Ham}
\begin{array}{l}
H=-\displaystyle\frac{1}{\kappa\sqrt{{\cal
E}E^xE^y}}\left[\rule{0mm}{8mm}\frac{1}{\gamma^2}\left\{ XE^xYE^y+{\cal
E}({\cal A}+\eta')(XE^x+YE^y)\right\}+\displaystyle\frac{1}{4}({\cal
E}')^2\right.+\\[4mm]
\left.\displaystyle\frac{1}{4}{\cal
E}^2\left(\frac{(E^x)'}{E^x}-\frac{(E^y)'}{E^y}\right)^2\right]
+\displaystyle\frac{1}{\kappa}\left(\displaystyle\frac{{\cal
EE}'}{\sqrt{{\cal
E}E^xE^y}}\right)'-\displaystyle\frac{\kappa\;G^2}{4\sqrt{{\cal
E}E^xE^y}}-\gamma\left(\sqrt{\frac{\cal E}{E^xE^y}}\,G\right)'.
\end{array}
\end{equation}
A prime denotes the derivative with respect
to $z$. As the last two terms of $H$ contain the Gau\ss\ constraint,
they may be dropped when the constraint is applied to
gauge-invariant states \footnote{This is not quite trivial, because
in quantum theory the inverse volume in these expressions, when
promoted to an operator, contains holonomies and so does not commute
with $G$. But, according to the usual factor ordering prescription
\cite{8}, holonomies stand left of triads, so that solutions of the
Gau\ss\ constraint are indeed annihilated by these parts of $H$.}.
When implemented these constraints are
integrated with test functions. For instance, the Hamiltonian
constraint is integrated with the lapse and denoted $H[N] =
\int dz N H$, as usual. With four canonical pairs of field variables and three first-class constraints
the system has one physical degree of freedom, which is realized by
polarized waves, moving in either direction along the $z$-axis.

We use two parts of the Hamiltonian constraint in subsequent
sections. The evolution of geometric quantities defined in the next
section requires the kinetic part $H_K$ of $H$, which contains the
conjugate variables $\cal A$, $X$, and $Y$, and is defined as in Ref. \cite{BD}
by the decomposition of the Hamiltonian constraint
\begin{equation}
H=-\frac{1}{\kappa}(H_K+H_P)
\end{equation}
with
\begin{equation}
H_K[N]=\frac{1}{\gamma^2}\int{\rm d}z\,N(z)\,\frac{XE^xYE^y+{\cal
E}({\cal A}+\eta')(XE^x+YE^y)}{\sqrt{{\cal E}E^xE^y}}(z).
\end{equation}
To construct the ``no wave" constraint we use the first term of
$H_K$,
\begin{equation}
H_K^1[N]=\frac{1}{\gamma^2}\int{\rm
d}z\,N(z)\,\frac{XE^xYE^y}{\sqrt{{\cal E}E^xE^y}}.
\end{equation}

Classically, we know that colliding waves produce a singularity
\cite{grif}. For this reason, and the fact that the goal is 
to investigate loop quantization and the dispersion
of gravitational waves, we further reduce the system to waves
propagating in only one direction \cite{HM2}. A space-time with
gravitational waves propagating exclusively in the negative
$z$-direction has a null Killing vector field in this direction,
related to uniform wave front propagation at the speed of light. As
shown in Ref. \cite{HM2}, the existence of this Killing vector field
yields a first-class ``left-moving constraint"
\begin{equation}\label{U+}
U_+:=XE^x+YE^y+\gamma{\cal E}'=0.
\end{equation}
Analogously, the first-class constraint
\begin{equation}\label{U-}
U_-:=XE^x+YE^y-\gamma{\cal E}'=0
\end{equation}
restricts to waves in the positive $z$ direction. Note that in Ref.
\cite{HM2} $U_-$ is defined as a {\em different} linear combination of
constraints that doesn't have the right-moving interpretation. The old
$U_-$ of  Ref. \cite{HM2} does not form a first-class algebra with
the Hamiltonian and diffeomorphism constraints, whereas the current
definition does. The Poisson brackets of the smeared-out constraint $U_+[f]=\int{\rm
d}z\,f(z)U_+(z)$ with test function $f$ and the other constraints
are weakly equal to zero.
\begin{eqnarray}
&&\{U_+[f],G[g]\}=0, \hspace{5mm}
\{U_+[f],C[g]\}= - U_+[f'g]\approx0,\\
&&\{U_+[f],H[N]\}= - U_+\left[\sqrt{\frac{\cal
E}{E^xE^y}}\,f' N\right]- \kappa \gamma H[ f N]\approx0.\label{sf}
\end{eqnarray}
Equation (\ref{sf}) contains the non-trivial structure function
$\sqrt{\frac{\cal E}{E^xE^y}}=\sqrt{g^{zz}}$, the square root of the
inverse metric component in the $z$-direction. Upon quantization,
this structure function becomes operator-valued and this may lead to
a quantum anomaly: The Dirac quantization procedure of determining
physical states by the condition that they be annihilated by the
constraint operators can be consistently carried out with equation
(\ref{sf}), when $U_+$ stands to the right of the structure
function. Otherwise new constraints may arise. For the full theory
it is shown in Ref. \cite{8} that a well-defined Hamiltonian
constraint is constructed from an operator ordering such that the
connection variables are left of the triad variables. 
For the spherically symmetric case, similar arguments are given in
Ref. \cite{2}. With such a factor ordering (or, also in symmetric
ordering) the first-class Poisson bracket relation (\ref{sf}) does
not obviously carry over to quantum theory without modification.

Before constructing plane-wave solutions, we construct flat space 
solutions as backgrounds for wave propagation. Flat space is modeled 
as a state without left- as well as right-moving waves. By imposing 
both left- and right-moving constraints we suppress all waves and 
arrive at a $(1+1)$-dimensional model without waves. Classically, 
imposing $U_+$ and $U_-$ simultaneously means
\begin{equation}\label{XEX}
{\cal K}:=XE^x+YE^y=0
\end{equation}
and
\begin{equation}\label{CalE}
{\cal E}'=0.
\end{equation}

These constraints, especially the first, will be the focus of the
rest of this work.  Together, the constraints form a first-class algebra
with the constraints of general relativity: They commute with the Gau\ss\
constraint. The Poisson brackets with the Hamiltonian constraint are
\begin{equation}
\{{\cal E}'[f],H[N]\}=-\frac{1}{\gamma}{\cal
K}\left[\sqrt{\frac{\cal E}{E^xE^y}}\,f'N\right], \hspace{5mm}
\{{\cal K}[f],H[N]\}=-\gamma{\cal E}'\left[\sqrt{\frac{\cal
E}{E^xE^y}}\,f'N\right]-\kappa\gamma H[fN],
\end{equation}
With the diffeomorphism constraint  the brackets are
\begin{equation}
\{{\cal K}[f],C[g]\}=-{\cal K}[f'g], \hspace{10mm} \{{\cal
E}'[f],C[g]\}=-{\cal E}'[f'g].
\end{equation}
Arising from the existence of a Killing vector field, we call
these pair of constraints the ``Killing constraints". (The singular
will refer to the first constraint, ${\cal K} =0$.) The second
constraint obviously expresses homogeneity in the $z$-direction. The
first constraint expresses homogeneity in the time direction, as we
will see in the next section. Additionally, the variables $X, Y, E^x,$ 
and $E^y$ span a subspace of the unconstrained phase space
of the total system.  The constraint ${\cal K}=0$, containing only these
variables, can be seen as a Hamiltonian function generating a flow
in this subspace along the vector field, $X \partial_X
+ Y \partial_Y - E^x \partial_{E^x} - E^y \partial_{E^y}$.  Among those
functions whose Lie derivatives vanish along this vector field are 
$XE^xYE^y$, which appears in the Hamiltonian constraint, and 
$E^x/E^y$, which is related to the ``wave factor" defined in \cite{MTW}.
The second ``no wave" constraint, ${\cal E}'=0$ containing a derivative, 
does not have such an interpretation.

\section{Geometric Quantities}
\label{GQ}

To develop a flat-space background geometry for plane waves it is
helpful to consider geometric quantities, as well as their time derivatives and interpretations,
with respect to co-moving observers, i.\,e. we
choose zero shift vectors and assume a fixed gauge of triads. This
kind of evolution is generated by the Hamiltonian constraint. The
time derivatives of the triad variables $\cal E$, $E^x$, and $E^y$
are given by Poisson brackets with $H$, in fact with the part
$-\frac{1}{\kappa}\,H_K$.

\noindent  {\bf Length}: A local measure of length is given by
\begin{equation}
\ell:=\sqrt{g_{zz}}=\sqrt{\frac{E^xE^y}{\cal E}}
\end{equation}
with the time derivative
\begin{equation}\label{dotl}
\dot\ell=\{\ell,H[N]\}=\frac{N}{\gamma}\,({\cal A}+\eta').
\end{equation}
A coordinate interval ${\cal I}$ has physical length
\begin{equation}\label{l}
\ell({\cal I})=\int_{\cal I}\sqrt{\frac{E^xE^y}{\cal
E}}\,{\rm d}z.
\end{equation}

\noindent{\bf Volume}: The local measure of volume is the square
root of the determinant of the spatial metric in equation (\ref{sm})
\begin{equation}
V=\sqrt{{\cal E}E^xE^y}.
\end{equation}
Its time derivative is
\begin{equation}
\dot V=\frac{N}{\gamma}\left(\frac{XE^x+YE^z}{\cal E}+{\cal
A}+\eta'\right).
\end{equation}

\noindent{\bf Cross section}: The geometrical meaning of the
quantity $\cal E$ is a cross section area, as $g_{xx}\cdot
g_{yy}={\cal E}^2$ is the determinant of the 2-metric in the $(x,z)$
plane. Its time derivative is
\begin{equation}\label{dotE}
\dot{\cal E}=\frac{N}{\gamma}\frac{XE^x+YE^y}{\ell}.
\end{equation}

\noindent The {\bf logarithmic ratio}, which is also called the ``wave factor" in Ref. \cite{MTW},
\begin{equation}
\beta=\ln\frac{E^x}{E^y}
\end{equation}
has the time derivative
\begin{equation}\label{dotb}
\dot\beta=\frac{N}{\gamma}\frac{YE^y-XE^x}{{\cal E}\ell}.
\end{equation}

\noindent{\bf Curvatures}: From Ref. \cite{HM2}, we have the
extrinsic curvature components ($K_a^i = e_b^i K_a^b$) in terms of
time derivatives
\begin{eqnarray}
K_x &:=&\sqrt{({K_x}^1)^2+({K_x}^2)^2}=\frac{1}{2N} \sqrt{ \frac{
E^x}{{\cal E} E^y} } \partial_t \left({\cal E}\frac{E^y}{E^x}\right)
\\[3mm]
K_y &:=& \sqrt{({K_y}^1)^2+({K_y}^2)^2}=\frac{1}{2N}
\sqrt{\frac{E^y}{{\cal E} E^x}} \partial_t \left({\cal
E}\frac{E^x}{E^y}\right)
\\[3mm]
K_z &\equiv& {K_z}^3 = \frac{1}{2N}\frac{\cal E}{E^xE^y} \partial_t
\left(\frac{E^xE^y}{\cal E}\right).
\end{eqnarray}
As we introduced in section \ref{variables}, $\gamma K_x=X$ and
$\gamma K_y=Y$. Using the Hamiltonian constraint to express the time
derivative we have
\begin{equation}
\label{K_z} K_z=  \frac{1}{\gamma} \sqrt{\frac{E^x E^y}{\mathcal
E}}(\mathcal{A} + \eta') = \frac{1}{\gamma} \frac{{\mathcal
E}}{V}(\mathcal{A} + \eta') = \partial_t \ln \ell.
\end{equation}

The above length and time derivatives of cross section area yield a
geometric interpretation of the Killing constraint $\cal K$. The
expression for the Killing constraint $\cal K$
\begin{equation}
\label{calK1}
{\cal K}=XE^x+YE^y=\frac{\gamma}{N}\,\ell\,\dot{\cal E}
\end{equation}
has the geometrical meaning of (length)$\times$(time derivative of
cross-section). We note that for a certain choice of the lapse
function, namely
\begin{equation}
N=\sqrt{\frac{E^xE^y}{\cal E}}=\sqrt{g_{zz}}=\ell,
\end{equation}
which is natural in the sense that it means $g_{tt}=g_{zz}$,
$XE^x+YE^y$ is precisely the time derivative of the cross section
area. With this choice inserted into (\ref{calK1}) we find ${\cal K}=\gamma \dot{\cal E}$. In $U_\pm$ this is set equal to $\mp$ $\cal E'$, which underlines the character of $U_\pm$ as plane wave constraints. Imposing ${\cal K}=0$ together with the constraint ${\cal E}'=0$, meaning that the area of the cross section in the $(x,y)$ plane is constant in $z$, enforces the
space-time translational invariance of the ``no wave"
state. This is the geometrical meaning of the Killing constraints,
equations (\ref{XEX}) and
(\ref{CalE}).\\

For an interpretation in terms of canonically conjugate variables it is
convenient to consider the triads as configuration variables and
carry out a canonical ``point transformation" to the new variables
$\cal E$, $\ell$, and $\beta$. To find new conjugate momenta,
corresponding to the time evolution introduced above, we first
express the variables $X$, $Y$, and $\cal A$ in terms of the time
derivatives $\dot\ell$, $\dot{\cal E}$, and $\dot\beta$ from
equations (\ref{dotl}), (\ref{dotE}), and (\ref{dotb}).

Using this we can construct the Lagrangian density
\begin{equation}
{\cal L}=\dot{\cal E}{\cal A}+\dot{E^x}X+\dot{E^y}Y-H.
\end{equation}
(Here we consider the Gau\ss\ constraint as satisfied, so that
$\eta$ is irrelevant.) The kinetic part of the Lagrangian is
\begin{equation}
L_K=\gamma\int\frac{{\rm
d}x}{N}\left(\frac{1}{4}\frac{\ell\,{\dot{\cal E}}^2}{\cal
E}-\frac{1}{4}\,{\cal E}\,\ell\,{\dot\beta}^2+\dot\ell\,\dot{\cal
E}\right).
\end{equation}
From this we can derive the conjugate momenta
\begin{eqnarray}
&&p_\ell=\frac{\delta L}{\delta\dot\ell
}=\frac{\gamma}{N}\,\dot{\cal E},\\
&&p_{\cal E}=\frac{\delta L}{\delta\dot{\cal
E}}=\frac{\gamma}{N}\left(\frac{\ell\,\dot{\cal E}}{2\,\cal
E}+\dot\ell\right),\\
&&p_\beta=\frac{\delta
L}{\delta\dot\beta}=-\frac{\gamma}{N}\,\frac{{\cal
E}\,\ell\,\dot\beta}{2}.
\end{eqnarray}
Using the time derivatives in terms of $X$, $Y$, and $\cal A$
gives the new momenta as functions of the original phase space
coordinates and so completes the
canonical transformation $(E^x,E^y,{\cal E};X,Y,{\cal
A})\leftrightarrow(\ell,{\cal E},\beta;p_\ell,p_{\cal E},p_\beta)$:
\begin{eqnarray}
&&p_\ell=\frac{1}{\ell}\,(XE^x+YE^y),\label{pl}\\
&&p_{\cal E}=\frac{XE^x+YE^y}{2\,\cal E}+{\cal A},\\
&&p_\beta=\frac{XE^x-YE^y}{2}.
\end{eqnarray}
From the first relation we see that the Killing constraint is the product of
length  and its conjugate momentum from equation (\ref{pl}),
\begin{equation}\label{kpl}
{\cal K}=\ell\, p_\ell,
\end{equation}
which will be of some interest when we discuss fluctuations.  Also note 
that due to the Poisson bracket relations $\{ {\cal K}, V \} = V$ and $\{ {\cal K}, \ell \} = \ell$ the 
constraint function $\cal K$ can also be interpreted as a volume and length dilatation generator.

\section{Quantum States and Operators}
\label{QSO}

In this section we briefly present quantum states and some basic
operators introduced in detail in Refs. \cite{BD,2}. We will then
apply the Killing constraints to this kinematic state space.

\subsection{Basic states}
Basic states are constructed from a one-dimensional version of spin
networks, denoted as ``charge-networks" in Ref. \cite{BD} with a graph
$G$ comprising edges and vertices along the $z$ axis. Along an edge
$e$ we define holonomies of the connection component $\cal A$,
\begin{equation}
h_e[{\cal A}]=\exp\left(i\frac{k_e}{2}\int_e{\cal A}\right).
\end{equation}
The edge label $k_e \in \mathbb{Z}$ denotes a representation of $U(1)$, so
the scalar density $\cal A$ appears in a natural way as a $U(1)$
connection. The connections
$X$ and $Y\in \mathbb{R}$ are scalars the natural holonomies of which are
point holonomies at the vertices $v$ (the location $z(v)$ of the
vertex $v$ will be frequently written as $v$)
\begin{equation}
h_v[X]=\exp\left(i\frac{\mu_v}{2}\,X(v)\right)
\hspace{5mm}\mbox{and}\hspace{5mm}
h_v[Y]=\exp\left(i\frac{\nu_v}{2}\,Y(v)\right)
\end{equation}
with vertex labels $\mu_v$ and $\nu_v$ in $\mathbb{R}$. These
holonomies are unitary representations of the Bohr compactification
of the reals, see Ref. \cite{BD,8}. The angular variable $\eta\in
\mathbb{R} / \mathbb{Z}$ gives rise to the point holonomies
\begin{equation}
h_v[\eta]=\exp(i\lambda_v\eta(v))
\end{equation}
in $U(1)$ with $\lambda_v\in \mathbb{Z}$. By application of the Gau\ss\
constraint these holonomies are expressed in terms of edge
holonomies and the labels $\lambda_v$ can be eliminated \cite{BD}.
A typical gauge-invariant state function based on a one-dimensional graph $G$ with
edges $e$ and vertices $v$ that is annihilated by the Gau\ss\
constraint is a product of the holonomies introduced above
\begin{equation}\label{T}
T_{G,\vec k,\vec\mu,\vec\nu}=\prod_{e\in
G}\exp\left[ i\frac{k_e}{2}\int_e\left\{ {\cal
A}(z)-\eta'(z)\right\} \right] \prod_{v\in
G}\exp\left(i\frac{\mu_v}{2}X\right)\exp\left(i\frac{\nu_v}{2}Y\right),
\end{equation}
These SNW functions, with all labels being nonzero, form an
orthogonal basis of the kinematical Hilbert space.

\subsection{Basic Operators}
The basic operators constructed from the configuration variables are
holonomy operators that act on state functions.
\begin{eqnarray}
&&\!\!\!\!\!\!\!\!\hat{h}_z({\cal I}):=\exp\left(\tau_3\int_{\cal
I}{\cal A}(z')\,{\rm d}z'\right)=\cos\left(\frac{1}{2}\int_{\cal
I}{\cal A}\right)+2\tau_3\sin\left(\frac{1}{2}\int_{\cal I}{\cal
A}\right),\label{hA}\\
&&\!\!\!\!\!\!\!\!\hat{h}_x(z):=\exp(\mu_0X(z)\tau_x(z))=
\cos\left(\frac{\mu_0}{2}X(z)\right)+2\tau_x(z)\sin\left(\frac{\mu_0}{2}X(z)\right),\label{hx}\\
&&\!\!\!\!\!\!\!\!\hat{h}_y(z):=\exp(\nu_0Y(z)\tau_y(z))=
\cos\left(\frac{\nu_0}{2}Y(z)\right)+2\tau_y(z)\sin\left(\frac{\nu_0}{2}Y(z)\right).\label{hy}
\end{eqnarray}
where $\cal I$ is some interval on the $z$ axis; $\mu_0$ and $\nu_0$
are parameters that determine the re\-presentation of the holonomy
to be created; the $k$-label of the edge holonomy created by
$\hat{h}_z$ is chosen to be equal to one. The matrices
$\tau_i=-i\sigma_i/2$ are $SU(2)$ generators. For the connection
$\cal A$ the generator $\tau_3$ is fixed, the $z$-dependent
generators $\tau_x$ and $\tau_y$ are defined by
\begin{equation}\label{tau}
\tau_x(z)=\cos \left( \eta(z) \right) \,\tau_1+\sin \left(\eta(z) \right) \,\tau_2,\hspace{5mm}
\tau_y(z)=-\sin \left( \eta(z) \right) \,\tau_1+\cos \left(\eta(z) \right)\,\tau_2.
\end{equation}
The conjugate variables give rise to flux operators. The scalar
${\cal E}(z)$ at an arbitrary point $z$ acts in the following way on
a state $T$
\begin{equation}\label{callE}
\hat{\cal E}(z)\,T_{G,\vec k,\vec\mu,\vec\nu}=\frac{\gamma\ell_{\rm
P}^2}{2} \frac{k_{+}(z)+k_{-}(z)}{2}\,T_{G,\vec k,\vec\mu,\vec\nu},
\end{equation}
where $k_\pm(z)$ denotes the edge labels on the two edges meeting at $z$,
if there is a vertex, or the edge label of one edge if there is no
vertex. (In this case $(k_++k_-)/2=k(z)$.) The fundamental length scale is set by $\ell_{\rm
P}^2=\kappa\hbar$.

The scalar densities $E^x$, $E^y$ have to be integrated over an
interval $\cal I$ to give the operators
\begin{eqnarray}
&&{\cal F}_x({\cal I})=\int_{\cal I}\hat{E}^x\,T_{G,\vec
k,\vec\mu,\vec\nu}= \frac{\gamma\ell_{\rm P}^2}{2}\sum_{v\in\cal
I}\mu_vT_{G,\vec
k,\vec\mu,\vec\nu},\\
&&{\cal F}_y({\cal I})=\int_{\cal I}\hat{E}^y\,T_{G,\vec
k,\vec\mu,\vec\nu}= \frac{\gamma\ell_{\rm P}^2}{2}\sum_{v\in\cal
I}\nu_vT_{G,\vec
k,\vec\mu,\vec\nu}.
\end{eqnarray}
Obviously, the flux operators are diagonal in the SNW basis.

\subsection{Geometric operators}
\label{geom_ops} The classical geometric quantities volume and
length may be quantized straightforwardly using LQG methods.\\

\noindent{\bf Volume}: Classically the volume of a block of space,
bounded by planes of unit coordinate area in the $x$ and $y$
directions, is, from equation (\ref{sm})
\begin{equation}
V({\cal I})=\int_{\cal I}{\rm d}z\,\sqrt{g}=\int_{\cal I}{\rm
d}z\sqrt{{\cal E}E^xE^y},
\end{equation}
over an interval $\cal I$ of coordinate length $\epsilon$. All the classical
triad variables are positive, $E^x$ and $E^y$ are radial variables,
and so $\cal E$ must be positive as long as the sign of the spatial
metric does not change. In quantum theory we allow for
both signs and take the absolute values in the volume. If $\cal I$
contains one vertex, we have
\begin{equation}
\begin{array}{ll}
V({\cal I})\approx\!\!&\!\epsilon\sqrt{|{\cal E}(v)E^x(v)E^y(v)|}=
\sqrt{|{\cal E}(v)|}\sqrt{\epsilon |E^x(v)|}\sqrt{\epsilon
|E^y(v)|}=\\[3mm]
&\sqrt{|{\cal E}(v)|}\sqrt{\left|\displaystyle\int_{\cal
I}E^x\right|}\sqrt{\left|\displaystyle\int_{\cal I}E^y\right|}.
\end{array}
\end{equation}
Inserting the corresponding flux operators and letting the resulting
volume operator $\hat V(\cal I)$ act on a vertex function of a SNW
state, defined by
\begin{equation}
\ket{v} := \,\ket{ k_\pm,\mu_v,\nu_v},
\end{equation}
gives $\hat{V}({\cal I}) \ket{v} =
V_v \ket{v}$ with the eigenvalue
\begin{equation}\label{43}
V_v=\frac{\gamma^\frac{3}{2}\ell_{\rm
P}^3}{4}\sqrt{|k_v||\mu_v||\nu_v|},
\end{equation}
where $k_v:=k_++k_-$ is the sum of the labels of the two adjacent
edges.\\

\noindent{\bf Length}: Analogously we may introduce a $z$-length
operator, starting from the classical length of an interval $\cal
I$. Unlike the volume, this expression for length contains $\cal E$
in the denominator. As the flux operator $\hat{\cal E}$ does not
have a densely defined inverse, we first replace the expression in
equation (\ref{l}) by applying Thiemann's identity \cite{8}
involving the Poisson bracket of quantities that have a direct
operator meaning. With the holonomy (\ref{hA}) we find, for small
intervals $\cal I$,
\begin{equation}
h_z({\cal I})\{h_z^{-1}({\cal
I}),V\}\approx-\gamma\kappa\,\frac{\tau_3}{2}\,\sqrt{\frac{E^xE^y}{\cal E}}.
\end{equation}
After quantization, when the Poisson bracket is replaced by
$(i\hbar)^{-1}$ times the commutator of the corresponding operators, we
conclude that the quantum operator of length can be written as
\begin{equation}\label{length}
\hat{\ell}({\cal I})={\rm
Tr}\left[\,\frac{-4i}{\gamma\ell_{\rm P}^2}\,\hat{h}_z({\cal
I})\,[\hat{h}_z^{-1}({\cal I}),\hat{V}({\cal I})]\tau_3\,\right].
\end{equation}
When applied to an interval $\cal I$ with one vertex $v$, $\hat\ell
({\cal I})$ gives the eigenvalue
\begin{equation}\label{47}
\ell_v =\frac{\sqrt{\gamma}\,\ell_{\rm
P}}{\sqrt{2}}\,\sqrt{|\mu_v||\nu_v|}\left(\sqrt{|k_v
+1|}-\sqrt{|k_v-1|}\right).
\end{equation}
When the edge labels $k_+$ and $k_-$ are large,
$\sqrt{k_v+1}-\sqrt{k_v-1}\approx1/\sqrt{k_v}$. So when the values
$\mu_v$ and $\nu_v$ at a vertex are fixed and the sum of the edge
labels is sent off toward infinity the length eigenvalues can become
arbitrarily small. In this limit $\ell_v$ becomes proportional to
$V_v/k_v$. This means that the length of a $z$-interval as
the thickness of a block of space in the $(x,y)$ plane is
approximately the block's volume divided by the $(x,y)$-area, given
by the eigenvalue of the flux operator $\hat{\cal E}$ (\ref{callE})
\footnote{The details of the calculation are carried out in the same
way as in the first part of the Hamiltonian constraint, with $\cal
I$ partitioned into a part left and a part right of the vertex,
see below. When $\cal I$ is taken as a whole, the expression in
parentheses in (\ref{47}) is
replaced by $\frac{1}{2}\left(\sqrt{|k_v+2|}-\sqrt{|k_v-2|}\right)$.}.\\

\noindent{\bf Inverse volume}: The quantization of the $V^{-1}$
operator proceeds by re-writing it in terms of the well-defined
classical quantities as done in \cite{BD}
\begin{equation}
V(I)^{-1} = \frac{16}{81\kappa^3\gamma^3
\mu_0\nu_0}\epsilon^{abc}{\rm Tr} \left[h_a\big\{h^{-1}_a,
V^{\frac{1}{3}}\big\}h_b\big\{h^{-1}_b,
V^{\frac{1}{3}}\big\}h_c\big\{h^{-1}_c, V^{\frac{1}{3}}\big\}
\right].
\end{equation}
So the quantum version is defined as
\begin{equation}
\widehat{V^{-1}} =  -\frac{16i}{81\hbar^3\kappa^3\gamma^3
\mu_0\nu_0}\epsilon^{abc}{\rm Tr}
\left[\hat{h}_a\left[\hat{h}^{-1}_a,
\hat{V}^{\frac{1}{3}}\right]\hat{h}_b\left[\hat{h}^{-1}_b,
\hat{V}^{\frac{1}{3}}\right]\hat{h}_c\left[\hat{h}^{-1}_c,
\hat{V}^{\frac{1}{3}}\right] \right]
\end{equation}
where $a,b,$ and $c$ are summed over
$x,y,z$. The action of this operator on a vertex is given in
Appendix \ref{inv_vol} with the result that
\begin{equation} \begin{split}
\widehat{V^{-1}} \ket{k_\pm,\mu_v,\nu_v} = V^{-1}_v \ket{k_\pm, \mu_v, \nu_v} \text{ with}\\
V^{-1}_v  = \frac{-1}{108 \ell_P^3 \gamma^{\frac{3}{2}}\mu_0\nu_0} \sqrt[3]{|k_v \mu_v\nu_v |}
\left(\sqrt[6]{| \mu_v-\mu_0|}-\sqrt[6]{|\mu_v+\mu_0|}\right) \\
\times \left(\sqrt[6]{|\nu_v-\nu_0|}-\sqrt[6]{|\nu_v+\nu_0|}\right)
\left(\sqrt[6]{|k_v -1|}-\sqrt[6]{| k_v +1| }\right).
\end{split} \end{equation}

\noindent{\bf Cross section}: Given the simple form of the cross section operator
its quantization is immediate.  At a vertex
\begin{equation}
\hat{{\cal E}} \ket{k_\pm, \mu_v, \nu_v} = \left( \frac{ \gamma
\ell_{\rm P}^2}{4} \right) k_v \ket{k_\pm, \mu_v, \nu_v}.
\end{equation}

\noindent{\bf Extrinsic Curvatures and zero curvature states}: Using
the holonomies of equations (\ref{hx},\ref{hy}), the quantization of
the $x$ extrinsic curvature is,
\begin{equation}
\widehat{K}_x := \frac{2}{\mu_0} {\rm Tr} \left[\tau_x (\hat{h}_x(z) - \hat{h}_x^{-1}(z) ) \right]
\end{equation}
The action at a single vertex is given by
\begin{equation}
\widehat{K}_x\left|k_\pm,\mu_v,\nu_v\right\rangle
=-\frac{i}{\kappa\gamma^2\mu_0}\left(\left|k_\pm,\mu_v+\mu_0,\nu_v\right\rangle
- \left|k_\pm,\mu_v-\mu_0,\nu_v\right\rangle \right).
\end{equation}
One could attempt to model flat space by requiring vertex states to
satisfy
\[
\widehat{K}_x\sum_{\mu_v} a_{\mu_v}\,\ket{k_\pm,\mu_v, \nu_v}=0.
\]
However this requires constant coefficients,
$a_{\mu_v+\mu_0} = a_{\mu_v-\mu_0}$, so such solutions are
non-normalizable and not in the kinematical Hilbert space.

The curvature operator in the $z$-direction has an ordering
ambiguity between the $z$ holo\-nomy, triad ${\cal E}$, and the
inverse volume operator.  However, there is only one Hermitian
ordering of these quantities.  For short intervals
${\cal I}$ we approximate
\[
({\cal A} + \eta') \approx 2 {\rm Tr} \left[ \tau_3( h_z( {\cal I}) - h^{-1}_z( {\cal I})) \right]
\]
and use this to define the quantum operator
\begin{equation}
\widehat{K}_z := 2 i \left( \hat{\cal E} \widehat{V^{-1}} \right)^{1/2} {\rm Tr} \left[ \tau_3( h_z( {\cal I})
 - h^{-1}_z( {\cal I})) \right] \left( \hat{\cal E} \widehat{V^{-1}} \right)^{1/2}
\end{equation}
The operator $\hat{K}_z$ has the action
\begin{equation} \begin{split}
\label{Kact}
\widehat{K}_z  \ket{  k_\pm, \mu_v, \nu_v,} &=
\mathcal{R}(\mu_v,\nu_v) {| k_v |}^{2/3} \left(\sqrt[6]{| k_v +1|}-
\sqrt[6]{| k_v-1| } \right)^{1/2} \\
& \ \ \ \times \bigg[ \left| k_v+1\right|^{2/3} \left(\sqrt[6]{ |
k_v +2|}-\sqrt[6]{| k_v| } \right) \ket{ k_\pm+1, \mu_v, \nu_v}
\\ & \ \ \ - \left| k_v-1\right|^{2/3} \left(\sqrt[6]{| k_v| }-\sqrt[6]{| k_v -2 |} \right) \ket{k_\pm-1, \mu_v, \nu_v}  \bigg]
 \end{split} \end{equation}
 where
\begin{equation}
 \mathcal{R}(\mu,\nu) := \frac{i}{648\mu_0\nu_0} \sqrt[3]{ | \mu\nu |}\left(\sqrt[6]{| \mu-\mu_0| }-\sqrt[6]{| \mu+\mu_0| }\right)
 \left(\sqrt[6]{| \nu-\nu_0| }-\sqrt[6]{| \nu+\nu_0|}\right)
\end{equation}
and $k_\pm+1$ ($k_\pm-1$) mean that both $k_+$ and $k_-$ are
raised (or lowered) by one on the intersection of the interval $\cal
I$ with the two adjacent edges of the vertex $v$.

For vanishing $z$ extrinsic curvature we consider (non-degenerate) states such that,
at every vertex,
\begin{equation}
\label{hardcon}
 \widehat{K}_z \sum_{k_v} a_{k_v} \ket{k_v,\mu_v, \nu_v} = 0
 \end{equation}
 Equation (\ref{Kact}) allows us to establish a recursion relation
between $a_{k_v+1}$ and $a_{k_v-1}$
\begin{equation}
\label{odb} a_{k_v+1} = \left[ \left(\frac{| k_v-1|}{|
k_v+1|}\right) \left(\frac{\sqrt[6]{| k_v |}-\sqrt[6]{|
k_v-2|}}{\sqrt[6]{| k_v+2 |}-\sqrt[6]{| k_v |}}\right) \right]^{1/2}
a_{k_v-1}
\end{equation}
This recursion relation iterates easily so that after $m$ terms
\begin{equation}
\label{biggie} a_{k_v+2m+1} = \left[ \left(\frac{ |k_v-1|}{|
k_v+2m+1|}\right) \left(\frac{\sqrt[6]{ |k_v| }-\sqrt[6]{|
k_v-2|}}{\sqrt[6]{| k_v+2m+2 |}-\sqrt[6]{| k_v+2m |}} \right)
\right]^{1/2} a_{k_v-1}
\end{equation}
For large $m$ then the coefficients scale as
\begin{equation}
a_{k+2m+1} \propto m^{- \frac{1}{12}}
\end{equation}
which does not converge fast enough to ensure normalization.

We see that
constraining any of the three extrinsic curvature operators to
vanish yields non-nomalizable states in the kinematic Hilbert space.
For this reason in the next section we will formulate flatness using
the Killing constraints which (eventually) yield normalizable solutions.

\subsection{The Hamiltonian Constraint Operator $\hat{H}_K^1$}

The formulation of the Killing constraint requires the
first part of the Hamiltonian constraint operator.
Our quantization is similar to Ref. \cite{2}, which is close to the
construction employed in full LQG, where the Hamiltonian constraint
is regularized in form of holonomies. Following this method we arrive
at a slightly different operator than in Ref. \cite{BD}.

The details are given in Appendix \ref{hk1} with the result that, on gauge invariant
states, $\ket{k_\pm,\mu_v,\nu_v}$,
\begin{equation}
\label{HK1}
\hat{H}_K^1[N] \ket{k_\pm,\mu_v,\nu_v}=\frac{1}{\sqrt{2}\gamma^{2}\mu_0\nu_0} \sum_v
N_v \ell_v \sin(\mu_0X)\sin(\nu_0Y) \ket{k_\pm,\mu_v,\nu_v}.
\end{equation}
(Up to a factor 2, this is equation (55) of Ref. \cite{BD}.) $N_v$
is the value of the lapse function at the vertex $v$. At each vertex
this term alters the labels $\mu_v$ and $\nu_v$ by $\pm\mu_0$ and
$\pm\nu_0$, respectively; it does not create new vertices. Whereas
in full theory it appears natural that the Hamiltonian constraint
changes the spin weights of edge holonomies by $\pm1/2$, there is no
{\em a priori} natural choice for $\mu_0$ and $\nu_0$ in the point
holonomies, which are in fact artifacts of the symmetry reduction.

We have seen that ``no-curvature" constraints yield non-normalizable states.
In the next sections we formulate and implement the Killing constraints.  This proves
to be not a simple matter of imposing the constraints, at least when we also ask that the
expectation values of length and volume on an atom of geometry be finite, but
requires a re-formulation of the Killing constraint.

\section{Formulating the Killing Constraint}
\label{KC}

Like the Hamiltonian constraint, the Killing constraint $\cal K$
in the form of equation (\ref{XEX}) contains connection variables
that do not have a direct meaning as operators. A substitute is
easily found in form of the Poisson bracket between the
well-defined volume operator and the first part $H_K^1$ of the
Hamiltonian constraint. Locally we have
\begin{equation}\label{Poi}
{\cal K}(z)=\frac{2}{\kappa\gamma} \left\{\frac{
XE^xYE^y}{\sqrt{{\cal E}E^xE^y}}(z),V({\cal
I})\right\}=2\,\frac{\gamma}{\kappa}\,\{H_K^1(z),V({\cal
I})\}.
\end{equation}
In consequence, the first version of the
Killing constraint $\hat{\cal K}$ can be defined as the
corresponding commutator
\begin{equation}
\hat{\cal K}:=i\,[\hat{H_K^1},\hat{V}({\cal I})].
\end{equation}
Note that we define the operator without the factor of 2.  Since the
action of the operator will vanish on states we also set $\gamma$
and the Planck length to 1 for the remainder of the paper.

The constraint turns out to act on each vertex individually
as $\hat{\cal K}|v\rangle=0$. Before obtaining solutions,
we note that for each vertex function $|v\rangle$ the
solutions, although being normalizable in the kinematic Hilbert
space, yields diverging expectation values for volume and length, as
will be shown in Section \ref{kill_c_ops}. For this reason we explore
modifications and generalizations of the above operator.

We can multiply $\cal K$ with an arbitrary positive power
of the volume, supposing the volume and length contribution of any
vertex are non-zero (justified later on in quantum
theory). Using the same algebra as equation (\ref{Poi}),
\begin{equation}\label{Vn}
V^{n-1}({\cal I}) {\cal
K}(z)=\frac{2}{n}\,\{H_K^1(z),V^n({\cal I})\},
\hspace{5mm} n\geq1,
\end{equation}
with $\cal K$ arising from $n=1$.

A similar modification of the Killing
constraint can be brought about by a modification of
$H_K^1$, denoted by
\begin{equation}
\label{Ha}
\left( H_K^1 \right)_p:= - 2 {\rm
Tr}\left[(h_xh_yh_x^{-1}h_y^{-1}-h_yh_xh_y^{-1}h_x^{-1})h_z\{h_z^{-1},V^p({\cal I})\}\right],
\end{equation}
which, to leading order, is
\begin{equation}\label{lHp}
p \, XE^xYE^y\,V^{p-2} =
p \, H_K^1\,V^{p-1}
\end{equation}
(The original $H_K^1$ is obtained by setting $p=1$.) The action
of the corresponding operator on a vertex state is
\begin{equation}\label{Halpha}
\hat{\left( H_K^1 \right)}_p \ket{v} =\frac{p}{2
\mu_0\nu_o}\left[(|\mu_v||\nu_v|)^\frac{p}{2}
\left(|k_v+2|^\frac{p}{2}-|k_v-2|^\frac{p}{2}\right)\sin(\mu_0X)\,\sin(\nu_0Y)\rule{0mm}{4mm}\right]
\ket{v}.
\end{equation}
The Poisson bracket of $V^q$ with (\ref{lHp}) (omitting the pre-factors) gives
\begin{equation}\label{ver}
\{V^q,\,XE^xYE^y\,V^{p-2}\}=\frac{q}{2}\,V^{p+q-2}\,(XE^x+YE^y).
\end{equation}
So for a given expression $V^n\cal K$ with $n=p+q-2$ there is a
two-parameter family of inequivalent commutators
$[\hat{\left( H_K^1 \right)}_p,\hat{V}^q]$ corresponding to
equation (\ref{ver}). They give rise to a two-parameter family of
modified operators, depending on $p$ and $q$, which will be denoted
by the ``volume weighted" Killing constraint
\begin{equation}
\label{Kpq} \hat{\cal K}_{p,q} := i [\hat{\left( H_K^1
\right)}_p,\hat{V}^q]
\end{equation}
with ${\cal K}_{1,1}=\cal K$. The meaning of these modifications
will be clear when we construct solutions to the Killing constraint.

\section{Implementing Quantum Killing Constraints}
\label{kill_c_ops}

The constraint ${\cal E}'=0$ is easy to handle as an operator. When
the scalar density ${\cal E}'$, respectively the operator density
$\hat{\cal E}'$, is integrated over an interval $\cal I$, we obtain
the flux operator (see equation (\ref{callE}))
\begin{equation}
\int_{\cal I}\hat{\cal E}'\,{\rm d}z=\hat{\cal E}_+-\hat{\cal E}_-,
\end{equation}
where $\hat{\cal E}_+$ and $\hat{\cal E}_-$ mean the operators
$\hat{\cal E}$ at the endpoints of the interval $\cal I$. Imposed as
a local constraint on SNW states, the solutions are simply states with
constant edge labels $k$.

The first version of the constraint $\cal K$ was represented as a
Poisson bracket in equation (\ref{Poi}).  This will be studied in
detail since this motivates the volume-weighted form of equation
(\ref{ver}). Integrating $\cal K$ over an interval $\cal I$ gives a
well-defined operator. When $\cal I$ contains a vertex, the action
of the corresponding operator is nontrivial.  With the sine
functions expanded, the operator has the action
\begin{eqnarray}
&&\hspace{-4mm} \hat{\cal K} | \mu_v,\nu_v\rangle =
\frac{1}{\mu_0\nu_0}\sqrt{|\mu_v||\nu_v||k|}
\left(\sqrt{|k+1|}-\sqrt{|k-1|}\right)\times\nonumber\\
&&\left\{\left(\sqrt{|\mu_v+2\mu_0||\nu_v+2\nu_0|}-\sqrt{|\mu_v||\nu_v|}\right)| \mu_v+2\mu_0,\nu_v+2\nu_0\rangle-\right.\nonumber\\
&&\left(\sqrt{|\mu_v-2\mu_0||\nu_v+2\nu_0|}-\sqrt{|\mu_v||\nu_v|}\right)| \mu_v-2\mu_0,\nu_v+2\nu_0\rangle-\\
&&\left(\sqrt{|\mu_v+2\mu_0||\nu_v-2\nu_0|}-\sqrt{|\mu_v||\nu_v|}\right)| \mu_v+2\mu_0,\nu_v-2\nu_0\rangle+\nonumber\\
&&\left.\left(\sqrt{|\mu_v-2\mu_0||\nu_v-2\nu_0|}-\sqrt{|\mu_v||\nu_v|}\right)| \mu_v-2\mu_0,\nu_v-2\nu_0\rangle\right\} .\nonumber
\end{eqnarray}
Since the edge labels $k$ are not changed we abbreviate the labels
in the remainder of this section and write simply
$|\mu_v,\nu_v\rangle$. Introducing $m_v:=\frac{\mu_v}{\mu_0}$ and
$n_v:=\frac{\nu_v}{\nu_0}$ we find
\begin{eqnarray}
&&\sqrt{|m_v||n_v||k|} \left(\sqrt{|k+1|}-\sqrt{|k-1|}\right)\times\nonumber\\
&&\left\{\left(\sqrt{|m_v+2||n_v+2|}-\sqrt{|m_v||n_v|}\right)| \mu_v+2\mu_0,\nu_v+2\nu_0\rangle-\right.\nonumber\\
&&\left(\sqrt{|m_v-2||n_v+2|}-\sqrt{|m_v||n_v|}\right)| \mu_v-2\mu_0,\nu_v+2\nu_0\rangle-\\
&&\left(\sqrt{|m_v+2||n_v-2|}-\sqrt{|m_v||n_v|}\right)| \mu_v+2\mu_0,\nu_v-2\nu_0\rangle+\nonumber\\
&&\left.\left(\sqrt{|m_v-2||n_v-2|}-\sqrt{|m_v||n_v|}\right)| \mu_v-2\mu_0,\nu_v-2\nu_0\rangle\right\}=0.\nonumber
\end{eqnarray}
Since this equation is identical for every vertex, we omit the index
$v$, and supposing that $\mu_v$ and $\nu_v$ are integer multiples of
$\mu_0$ and $\nu_0$, respectively, we write the vertex state as
\begin{equation}\label{81}
|v \rangle=\sum_{m,n}a_{m,n}| m\mu_0,n\nu_0\rangle
\end{equation}
with coefficients $a_{m,n}$.

At each vertex the following difference
equation arises from imposing the Killing constraint
\begin{eqnarray}
\label{76}
&&\sqrt{|m-2||n-2|}\left(\sqrt{|m||n|}-\sqrt{|m-2||n-2|}\right)\,a_{m-2,n-2}-\\
&&\sqrt{|m+2||n-2|}\left(\sqrt{|m||n|}-\sqrt{|m+2||n-2|}\right)\,a_{m+2,n-2}-\nonumber\\
&&\sqrt{|m-2||n+2|}\left(\sqrt{|m||n|}-\sqrt{|m-2||n+2|}\right)\,a_{m-2,n+2}+\nonumber\\
&&\sqrt{|m+2||n+2|}\left(\sqrt{|m||n|}-\sqrt{|m+2||n+2|}\right)\,a_{m+2,n+2}=0\nonumber
\end{eqnarray}
This equation applies to the (rescaled) vertex labels $m$ and $n$
and connects quadruples of coefficients $a_{m,n}$. Due to the
constant nature of the edge labels, $k$ we have omitted the factors
depending on these labels.

Before considering solutions, we observe that due to the occurrence
of absolute values in the length and volume eigenvalues there is
some degeneracy: they do not depend on the signs of $\mu_v$ and
$\nu_v$, and the signs of $m$ and $n$. Thus, we can characterize
an ``orientation" of basic states: $(+)$ if ${\rm sgn}(\mu_v)={\rm
sgn}(\nu_v)$ and $(-)$ if ${\rm sgn}(\mu_v)=-{\rm sgn}(\nu_v)$. (A
similar orientation plays a role in loop quantum cosmology
\cite{cos}.) As we will see below, it is possible to construct
states with pure $(+)$ or $(-)$ orientation. In the following we
concentrate on positively oriented states. (The question of whether
orientation is conserved under the evolution generated by the
Hamiltonian constraint will be left to future work.)

To exhibit {\em a} solution of $a_{m,n}$ to the Killing constraint we begin with
inserting $m=n=0$ into equation (\ref{76}), which leads to
\begin{equation}
a_{-2,-2}-a_{2,-2}-a_{-2,2}+a_{2,2}=0.
\end{equation}
Two of the coefficients can be chosen to be zero. We choose
$a_{-2,2}=a_{2,-2}=0$ and find
\begin{equation}
a_{-2,-2}=-a_{2,2}.
\end{equation}
For $m=4$, $n=0$ we have
\begin{equation}
a_{2,-2}-3a_{6,-2}-a_{2,2}+3a_{6,2}=0
\end{equation}
and, with the choice $a_{6,-2}=0$ we get
\begin{equation}
a_{6,2}=\frac{1}{3}\,a_{2,2}.
\end{equation}
Continuing in this way by setting $a_{m,-2}=0$ for $m>0$, we obtain
the nonzero coefficients
\begin{equation}
a_{4r+2,2}=\frac{1}{2r+1}\,a_{2,2}
\end{equation}
for $r\geq0$.

Setting also $a_{m,2}=0$ for $m<0$ and $a_{2,n}=0$ for $n<0$, 
we obtain a solution with nonzero coefficients confined to the
first and third quadrants in the $(m,n)$ plane. This solution has
the symmetry properties $a_{-m,-n}=-a_{m,n}$ and $a_{m,n}=a_{n,m}$.
It has pure $(+)$ orientation and avoids zero volume or zero length
states with $m=0$ or $n=0$. It may be characterized by one
fundamental initial value $a_{2,2}$.
Under the assumption of integer $m$ and $n$ and
positive orientation there are eight linearly independent
solutions with fundamental initial values $a_{1,1}$, $a_{1,2}$,
$a_{1,3}$, $a_{2,1}$, $a_{3,1}$, $a_{2,2}$, $a_{2,3}$, and
$a_{3,2}$.

If we admit non-integer values of $m$ and $n$, we can construct
analogous oriented solutions with fundamental initial value
$a_{m,n}$ with $0<m<2$ and $0<n<2$. Non-integer values of $m$ and
$n$ mean that the vertex labels $\mu_v$ and $\nu_v$ are not integer
multiples of the labels $\mu_0$ and $\nu_0$ in the Hamiltonian
constraint. Supposing they are integer multiples renders the volume
and length spectra discrete, such an assumption has the advantage to
reflect the discreteness of full LQG in our 1+1 dimensional model.

Numerical calculations of this solution indicate that they are normalizable in the kinematical
Hilbert space, but the volume
and length expectation values (at each vertex),
\begin{equation}
\langle V_v \rangle\propto\sum_{m,n} |a_{m,n}|^2 \sqrt{|k||m||n|}
\end{equation}
and
\begin{equation}
\langle \ell_v
\rangle\propto\sum_{m,n}|a_{m,n}|^2\sqrt{|m||n|}\left(\sqrt{|k+1|}-\sqrt{|k-1|}\right)
\end{equation}
(see equations (\ref{43}) and (\ref{47})), diverge.

To investigate the asymptotic behavior of the solutions for large
$m$ and $n$, we approximate the coefficients by a continuous
function $a(m,n)$ and the difference equation (\ref{76}) by a
differential equation
\begin{equation}
2m\, \partial_m a +2n\, \partial_n a +3a=0.
\end{equation}
A simple solution is
\begin{equation}
a(m,n)=m^\alpha\,n^\beta \hspace{5mm} \mbox{where} \hspace{5mm}
\alpha+\beta=-\frac{3}{2}.
\end{equation}
However, it is not hard to show that ensuring finite expectation value of length or volume,
\[ \sum_{m,n} \sqrt{mn}\left|a(m,n)\right|^2 < \infty, \]
requires that $\alpha+\beta< -\tfrac{3}{2}$. Thus finite expectation values are not possible with
the first version of the Killing constraint. 

The divergence of the volume and length expectation values in a
state on which the Killing constraints are exactly satisfied,
indicates that such conditions are too strong for a physical quantum
state. Already on the kinematical level we see that we cannot solve
the Killing constraint equation $\hat{\cal K}|v\rangle=0$ for a
kinematical vertex state $|v\rangle$ with reasonable physical
properties. This is why we explore modifications.

Recall from equation (\ref{kpl}) that $\cal K$ can be represented as
the product of two canonically conjugate variables. In
one-dimensional quantum mechanics wave functions on the interval
$(0,\infty)$ that are annihilated by $\hat q\hat p$, $\hat p\hat q$
or the anticommutator $[\hat q,\hat p]_+$ are non-normalizable, so
the problems with solutions of $\hat{\cal K}|v\rangle=0$ may not
come as a big surprise.

As discussed in the last section, it is possible to weaken
the Killing constraint equation in order to find physically
acceptable solutions. The volume-weighted Killing constraint operator $\hat{\cal
K}_{pq}$ of equation (\ref{Kpq}) leads to difference equations of
the form
\begin{eqnarray}\label{difq}
&&(|m-2||n-2|)^\frac{p}{2}\left((|m||n|)^\frac{q}{2}-(|m-2||n-2|)^\frac{q}{2}\right)\,a_{m-2,n-2}-\\
&&(|m+2||n-2|)^\frac{p}{2}\left((|m||n|)^\frac{q}{2}-(|m+2||n-2|)^\frac{q}{2}\right)\,a_{m+2,n-2}-\nonumber\\
&&(|m-2||n+2|)^\frac{p}{2}\left((|m||n|)^\frac{q}{2}-(|m-2||n+2|)^\frac{q}{2}\right)\,a_{m-2,n+2}+\nonumber\\
&&(|m+2||n+2|)^\frac{p}{2}\left((|m||n|)^\frac{q}{2}-(|m+2||n+2|)^\frac{q}{2}\right)\,a_{m+2,n+2}=0\nonumber
\end{eqnarray}
the solutions of which fall off more rapidly with growing $m$ and
$n$, when $p$ and $q$ are large enough.

We consider states created as solutions of the generalized
difference equation (\ref{difq}), first with $p+q=2$, so that all
versions of the constraint operators correspond to the same
classical expression $\cal K$ of equation (\ref{Poi}) (and the 
right-hand side of equation (\ref{ver}) without the factor of volume). 
In the next step this condition is relaxed and cases
with $p+q>2$ are considered. Concretely, we estimate three
quantities: The average value
\begin{equation}
\langle
W\rangle=\left\langle\sqrt{\left|\frac{\mu_v}{\mu_0}\right|\cdot\left|\frac{\nu_v}{\nu_0}\right|}\right\rangle=\langle\sqrt{|m||n|}\rangle,
\end{equation}
which, for a fixed value of the edge parameter $k$, is proportional
to both the length and the volume expectation values, the
uncertainty
\begin{equation}
\Delta W=\left(\langle W^2\rangle-\langle
W\rangle^2\right)^\frac{1}{2},
\end{equation}
and the departure of a considered state function from being
annihilated by $\hat{\cal K}$. Since the Killing constraint operator
$\hat{\cal K}$ is not Hermitian we calculate, from the original
Killing constraint, the moment
\begin{equation}
\label{uncert_K}  ||\hat{\cal K} |v\rangle || := 
\langle v|\hat{\cal K}^\dag\hat{\cal K} |v\rangle^\frac{1}{2}
\end{equation}
instead of $\langle\hat{\cal K}^2\rangle^\frac{1}{2}$, which
contains positive and negative contributions.

We numerically calculated the quantities by inserting the
coefficients into an $(m,n)$ diagram and forming sums over diagonals
with slope $-1$ in the first quadrant, that is, e.g. $a_{2,2}$,
$a_{2,4}+a_{4,2}$, $a_{2,6}+a_{4,4}+a_{6,2}$, and so on. The
contributions of each diagonal can be fairly well fitted by a
decreasing power function, and after summing up 10-15 of them, the
remainders were estimated by integrals over the extrapolated power
functions. The contributions of these integrals to the estimates of
the infinite sums is of the order of a few percent.

For the original constraint $\cal K$ with $p=q=1$ the expectation
value $\langle W\rangle$ diverges. Further, in all cases $p+q=2$ the
uncertainties of length and volume diverge, and for $p>q$ they diverge more rapidly. For
instance, with $p=3/2$, $q=1/2$ we have
\begin{equation}
\langle W\rangle=5.41,\hspace{1cm}\Delta
W=\infty,\hspace{1cm} ||\hat{\cal K}|v\rangle|| =1.05.
\end{equation}
In fact as the parameters depart further
from the ``natural" values of $p=q=1$ we obtain decreasing
expectation values, but $\Delta W$ always diverges. This need not
necessarily disqualify a state as a physical state, but in a
realistic quantum model for flat space such states can have only a
tiny or zero probability measure. For $p<q$ the divergence of
$\langle W\rangle$ becomes worse than for $p=q=1$.

For $p+q\geq2$ some results are summarized in Tables I-III. From the
Table I we see that with growing $p+q$ the average value $\langle
W\rangle$ goes quickly to 2, which means that $\ket{2, 2}$ becomes
dominant. Table II confirms this tendency. The limiting state for
$p+q$ growing to infinity is the basis state $| 2,2\rangle$, i.\,e.
all coefficients vanish except $a_{2,2}$, which is equal to one. In
this state $||\hat{\cal K}| 2,2\rangle||=4\sqrt{3}\approx6.92$,
which is an upper bound on Table III and so the maximal departure of
the considered states from $\hat{\cal K}|v\rangle=0$.\\[3mm]
\parbox{6cm}{
\hspace{5.5mm}\begin{tabular}{c||c|c|c|c}
\raisebox{-1mm}{$q$}$\backslash$\raisebox{1mm}{$p$} &
\raisebox{-1mm}{1} & \raisebox{-1mm}{2} & \raisebox{-1mm}{3} &
\raisebox{-1mm}{4} \\[1mm]
\hline\hline
\raisebox{-1mm}{1} & \raisebox{-1mm}{$\infty$} & \raisebox{-1mm}{2.25} & \raisebox{-1mm}{2.052} & \raisebox{-1mm}{2.014} \\[1mm]
\hline
\raisebox{-1mm}{2} & \raisebox{-1mm}{2.55} & \raisebox{-1mm}{2.065} & \raisebox{-1mm}{2.015} & \raisebox{-1mm}{2.0045} \\[1mm]
\hline
\raisebox{-1mm}{3} & \raisebox{-1mm}{2.13} & \raisebox{-1mm}{2.02} & \raisebox{-1mm}{2.005} & \raisebox{-1mm}{2.0015}\\[1mm]
\hline \raisebox{-1mm}{4} & \raisebox{-1mm}{2.04} &
\raisebox{-1mm}{2.007} & \raisebox{-1mm}{2.002} &
\raisebox{-1mm}{2.0005}
\end{tabular}
} \ \hspace{1.55cm} \ \parbox{8cm}{
TABLE I: The expectation value $\langle W\rangle$ as a function of
the para\-meters $q$ and $p$. The expectation value peaks on the
$\ket{2,2}$ state as the parameters increase.}
\vspace{1cm}

\parbox{6cm}{
\begin{tabular}{c||c|c|c|c}
\raisebox{-1mm}{$q$}$\backslash$\raisebox{1mm}{$p$} &
\raisebox{-1mm}{1} & \raisebox{-1mm}{2} & \raisebox{-1mm}{3} &
\raisebox{-1mm}{4} \\[1mm]
\hline\hline
\raisebox{-1mm}{1} & \raisebox{-1mm}{$\infty$} & \raisebox{-1mm}{0.82} & \raisebox{-1mm}{0.112} & \raisebox{-1mm}{0.108} \\[1mm]
\hline
\raisebox{-1mm}{2} & \raisebox{-1mm}{$\infty$} & \raisebox{-1mm}{0.236} & \raisebox{-1mm}{0.087} & \raisebox{-1mm}{0.065} \\[1mm]
\hline
\raisebox{-1mm}{3} & \raisebox{-1mm}{0.75} & \raisebox{-1mm}{0.103} & \raisebox{-1mm}{0.052} & \raisebox{-1mm}{0.038}\\[1mm]
\hline \raisebox{-1mm}{4} & \raisebox{-1mm}{0.33} &
\raisebox{-1mm}{0.060} & \raisebox{-1mm}{0.029} &
\raisebox{-1mm}{0.022}
\end{tabular}
} \hspace{1.3cm}
\parbox{8cm}{
TABLE II: The uncertainty $\Delta W$ as a function of the parameters
$q$ and $p$. The uncertainty decreases rapidly with increasing parameters.}

\vspace{1cm}

\parbox{6cm}{
\begin{tabular}{c||c|c|c|c}
\raisebox{-1mm}{$q$}$\backslash$\raisebox{1mm}{$p$} &
\raisebox{-1mm}{1} & \raisebox{-1mm}{2} & \raisebox{-1mm}{3} &
\raisebox{-1mm}{4} \\[1mm]
\hline\hline
\raisebox{-1mm}{1} & \raisebox{-1mm}{0} & \raisebox{-1mm}{3.51} & \raisebox{-1mm}{5.078} & \raisebox{-1mm}{5.88} \\[1mm]
\hline
\raisebox{-1mm}{2} & \raisebox{-1mm}{2.86} & \raisebox{-1mm}{4.87} & \raisebox{-1mm}{5.82} & \raisebox{-1mm}{6.31} \\[1mm]
\hline
\raisebox{-1mm}{3} & \raisebox{-1mm}{4.51} & \raisebox{-1mm}{5.70} & \raisebox{-1mm}{6.27} & \raisebox{-1mm}{6.57}\\[1mm]
\hline \raisebox{-1mm}{4} & \raisebox{-1mm}{5.47} &
\raisebox{-1mm}{6.19} & \raisebox{-1mm}{6.54} &
\raisebox{-1mm}{6.72}
\end{tabular}}
\hspace{1.4cm}
\parbox{8cm}{TABLE III: The moment $||\hat{\cal K} |v\rangle ||$ defined in equation (\ref{uncert_K}), as
a function of the parameters $q$ and $p$.  It is finite and bounded
by $||\hat{\cal K}| 2,2\rangle||=4\sqrt{3}\approx6.92$.}
\vspace{1cm}

The volume-weighted Killing constraint does a better job of
approximating the flat space limit of the plane gravitational
space-time in that it has finite geometric expectation values and
decreasing uncertainties (with $q$ and $p$). However, 
the non-vanishing moment $||\hat{\cal K}|v\rangle||$ and the non-hermiticity
of the original Killing constraint leads us to consider a different
formulation of the Killing constraint, which is described in the
next section. In this case we can solve the constraint and perform a
similar analysis of the uncertainties.

\section{A Hermitian Killing Constraint Operator}
\label{HKC}

As an alternative to the $\hat{\cal K}_{p,q}$ operator in the two
foregoing sections a quite simple, Hermitian Killing
operator can be constructed at least in the weak field limit. The
construction arises from an approximation of $X$ and $Y$, which is
valid classically for small $X$ and $Y$, i.\,e. for a weak
gravitational field, and is an alternative to the Thiemann trick.
We may approximate
\begin{equation}\label{X}
X(z)\approx\frac{2}{\mu_0}\,{\rm
Tr}\left[\tau_x h_x^{-1}(z)\right]\approx\frac{2}{\mu_0}\,{\rm
Tr}\left[\tau_x+\frac{1}{4}\,\mu_0X(z)\right],
\end{equation}
and analogously $Y$. For edge holonomies such an approximation
becomes exact in the continuous limit, when edges grow arbitrarily
short. For point holonomies, which are an artifact of the
homogeneity in the $x$ and $y$ directions, we must assume
$X\ll\frac{4}{\mu_0}$ in order to replace $X$ by the right-hand side
of equation (\ref{X}).

\subsection{Modified constraint operator and equation}
With the aid of equation (\ref{X}), the modified Killing constraint
\begin{equation}\label{98}
{\cal K}_q:=V^q{\cal K}V^q
\end{equation}
can be arranged in the following symmetric operator form
\begin{equation}\label{3}
\hat{\cal K}_q[{\cal I}]=2\int_{\cal I}{\rm
d}z\,\hat{V}^q\left(\sqrt{{\hat E}^x}\,\frac{{\rm
Tr}\left[\tau_x\hat{h}^{-1}_x[{\cal I}]\right]}{\mu_0}\,\sqrt{{\hat
E}^x}+\sqrt{{\hat E}^y}\,\frac{{\rm
Tr}\left[\tau_y\hat{h}^{-1}_y[{\cal I}]\right]}{\nu_0}\,\sqrt{{\hat
E}^y}\right){\hat V}^q.
\end{equation}
As before, we seek a superposition of vertex states such that
\begin{equation}\label{8}
\hat{\cal K}_q\,\sum_{\mu,\nu}a_{\mu,\nu}|k,\mu,\nu\rangle=0,
\end{equation}
where the label $v$ of $\mu$ and $\nu$ has been suppressed. The
analysis is similar to Section \ref{kill_c_ops} and the details are
presented in Appendix \ref{HKCsol}. It is also shown there that,
like the previous formulation of the Killing constraint, this
operator must be volume-weighted in order to ensure finite
expectation values of length and volume. In Appendix C it is also
shown that $\mu$ and $\nu$ in (\ref{8}) must not be even integer
multiples of $\mu_0$ and $\nu_0$. For odd integer multiples the
resulting normalized states are
\begin{equation}
\label{q_solns}
|q\rangle=\frac{2^q}{2^{q+1}-1}\frac{1}{\zeta(q+1)}\sum_{m,n=-\infty}^\infty\frac{1}{|(2m+1)(2n+1)|^\frac{q+1}{2}}\,
|2m+1,2n+1\rangle,
\end{equation}
in the notation 
\begin{equation}
\label{statenotation}
|2m+1,2n+1\rangle:=|k,(2m+1)\mu_0,(2n+1)\nu_0\rangle,
\end{equation} and where $\zeta(q)$ is the
Riemann zeta function. These states have finite expectation values of length and volume only when $q>1$.

In the limit of large $q$ the state $|q\rangle$ reduces to a
superposition of four states with the same length and volume
expectation values,
\begin{equation}\label{29}
\lim_{q\rightarrow\infty}|q\rangle\longrightarrow\frac{1}{2}\left(|1,1\rangle+|1,-1\rangle+|-1,1\rangle+|-1,-1\rangle\right),
\end{equation}
and the uncertainties for length and volume go to zero, as we can see from
\begin{equation}
\label{HKCDW}
\Delta
W=2\,\frac{[(2^q-1)^2(2^{q+1}-1)^2\zeta(q)^2\zeta(q+1)^2-(2^{q+\frac{1}{2}}-1)^4\zeta(q+\frac{1}{2})^4]^\frac{1}{2}}
{(2^{q+1}-1)^2\,\zeta(q+1)^2}.
\end{equation}

\subsection{Uncertainty in the Killing Constraint}
As in the previous section we consider the departure of our
solution states from being annihilated by the original Killing
constraint. The constraint $\hat{\cal K}_q$ is now Hermitian so it makes sense to
calculate the usual uncertainty $\left(\langle\hat{\cal
K}^2\rangle-\langle\hat{\cal K}\rangle^2\right)^\frac{1}{2}$, where
$\langle\hat{\cal K}\rangle$ is again equal to zero. In the
following we consider a slight generalization, namely the
uncertainty
\begin{equation}
\Delta{\cal K}_p=\langle q|\,\hat{\cal
K}_p^2\,|q\rangle^\frac{1}{2},
\end{equation}
where the parameter of the constraint (\ref{98}) was denoted by $p$
and may be different from the state label $q$. The expression of the
constraint is the same, we have only weighted the state and the
uncertainty operator differently. This will allow us to perform
three very similar calculations at once.

The operator $\hat{\cal K}_p$ acts on a vertex state $|k,\mu,\nu\rangle$ by replacing it
by four states in the neighborhood, namely
$|k,\mu\pm\mu_0,\nu\pm\nu_0\rangle$. By acting with $\hat{\cal K}_p$
on $|q\rangle$  in the notation of (\ref{statenotation}) we obtain from
$|2m+1,2n+1\rangle$ and its neighboring states
$|2(m\pm1)+1,2(n\pm1)+1\rangle$ contributions to the states
$$|2m,2n+1\rangle,\hspace{5mm}|2m+2,2n+1\rangle,\hspace{5mm}
|2m+1,2n\rangle,\hspace{5mm} |2m+1,2n+2\rangle.
$$
Let's consider $|2m,2n+1\rangle$ in detail. From the action of
$\hat{\cal K}_p$ on $|2m+1,2n+1\rangle$ we find
$$i\left(\frac{1}{4^{2p}}\,2^\frac{p+1}{2}\,|k|^p(\mu_0\nu_0)^p\right)|2n+1|^p\left(|m||2m+1|\right)^\frac{p+1}{2}
|2m,2n+1\rangle,$$ 
and from the action on $|2m-1,2n+1\rangle$ we find
$$-i\left(\frac{1}{4^{2p}}\,2^\frac{p+1}{2}\,|k|^p(\mu_0\nu_0)^p\right)|2n+1|^p(|m||2m-1|)^\frac{p+1}{2}|2m,2n+1\rangle.
$$
The prefactor in parenthesis, which is the same in all cases, can be
omitted because we are primarily interested in numerical comparisons for
different parameters $q$ and $p$. This means that we are effectively
studying the single vertex behavior of the states.
Up to this prefactor, $\hat{\cal K}_p$ generates the state
\begin{equation}
i|2n+1|^p\beta_{2n+1}|m|^\frac{p+1}{2}(|2m+1|^\frac{p+1}{2}\alpha_{2m+1}-|2m-1|^\frac{p+1}{2}\alpha_{2m-1})|2m,2n+1\rangle.
\end{equation}
where coefficients $\alpha$ and $\beta$ are solutions to the separated difference equations (\ref{alpha_eqn}) and (\ref{beta_eqn}). After insertion of these coefficients this becomes
\begin{equation}\label{37}
i|2n+1|^{p-\frac{q+1}{2}}|m|^\frac{p+1}{2}(|2m+1|^\frac{p-q}{2}-|2m-1|^\frac{p-q}{2})|2m,2n+1\rangle.
\end{equation}
The other three states created from $|2m+1,2n+1\rangle$ by
$\hat{\cal K}_p$ are
\begin{eqnarray}
&&\hspace{-15mm}i|2n+1|^{p-\frac{q+1}{2}}|m+1|^\frac{p+1}{2}(|2m+3|^\frac{p-q}{2}-|2m+1|^\frac{p-q}{2})|2m+2,2n+1\rangle,\label{38}\\[2mm]
&&\hspace{-15mm}i|2m+1|^{p-\frac{q+1}{2}}|n|^\frac{p+1}{2}(|2n+1|^\frac{p-q}{2}-|2n-1|^\frac{p-q}{2})|2m+1,2n\rangle,\label{39}\\[2mm]
&&\hspace{-15mm}i|2m+1|^{p-\frac{q+1}{2}}|n+1|^\frac{p+1}{2}(|2n+3|^\frac{p-q}{2}-|2n+1|^\frac{p-q}{2})|2m+2,2n+2\rangle.\label{40}
\end{eqnarray}

A second application of $\hat{\cal K}_p$ on $|q\rangle$ shifts the
contributions back to the ``old" places at odd multiples of $\mu_0$
and $\nu_0$ and to ``new" places with both multiples of $\mu_0$ and
$\nu_0$ being even, but only the former ones contribute to $\langle
q|\hat{\cal K}_p|q\rangle$. In particular, the action of $\hat{\cal
K}_p$ on the four states (\ref{37},\ref{38},\ref{39},\ref{40})
contribute to the state $|2m+1,2n+1\rangle$, and multiplying this
contribution by $a_{2m+1,2n+1}$ we get (up to an overall factor) the
matrix element
\begin{eqnarray}\label{112}
&&\hspace{-5mm}\langle 2m+1,2n+1|\,\hat{\cal K}_p^2\,|2m+1,2n+1\rangle \propto \nonumber\\[2mm]
&&\hspace{5mm}|2n+1|^{2p-q-1}|2m+1|^\frac{p-q}{2}\left[|m|^{p+1}\left(|2m+1|^\frac{p-q}{2}-|2m-1|^\frac{p-q}{2}\right)\right.\nonumber\\[1mm]
&&\hspace{5mm}\left.-|m+1|^{p+1}\left(|2m+3|^\frac{p-q}{2}-|2m+1|^\frac{p-q}{2}\right)\right]\\[1mm]
&&\hspace{5mm}+|2m+1|^{2p-q-1}|2n+1|^\frac{p-q}{2}\left[|n|^{p+1}\left(|2n+1|^\frac{p-q}{2}-|2n-1|^\frac{p-q}{2}\right)\right.\nonumber\\[1mm]
&&\hspace{5mm}\left.-|n+1|^{p+1}\left(|2n+3|^\frac{p-q}{2}-|2n+1|^\frac{p-q}{2}\right)\right].\nonumber
\end{eqnarray}

In the following, the values $p=0$ and $p=-1/2$ will be of interest.
The first value corresponds to the original, un-weighted Killing
constraint $\cal K$ while the second value gives an expression that
contains a factor of the momentum $\hat{p}_\ell$ at a single vertex
\footnote{More precisely, the operator ${\cal K}_{-1/2} = p_\ell/
{\cal E}$. We have neglected the contribution of ${\cal E}$ due to
the first Killing constraint and the constancy of $k_v$.}. Although
the inverse length in the second one cannot be directly formulated
as a densely defined operator, the above formula makes sense also
for $p=-1/2$. This means that formula (\ref{112}) can be
extrapolated for $p=-1/2$ for states of the form $|q\rangle$ with
$q>1$.

For $p=0$ and $q=2$ we have matrix elements
\begin{eqnarray}
&&\hspace{-1cm}\langle2m+1,2n+1|\,\hat{\cal
K}^2\,|2m+1,2n+1\rangle \propto \nonumber\\[2mm]
&&|2n+1|^{-3}|2m+1|^{-1}\left[|m|\left(|2m+1|^{-1}-|2m-1|^{-1}\right)\right.\nonumber\\
&&-\left.|m+1|\left(|2m+3|^{-1}-|2m+1|^{-1}\right)\right]\\
&&+|2m+1|^{-3}|2n+1|^{-1}\left[|n|\left(|2n+1|^{-1}-|2n-1|^{-1}\right)\right.\nonumber\\
&&-|n+1|\left.\left(|2n+3|^{-1}-|2n+1|^{-1}\right)\right],\nonumber
\end{eqnarray}
their sum is numerically approximately equal to 1.052; for $q=4$ we
find 1.610. In the limit $q\rightarrow\infty$ only $m=n=0$
contributes and
\begin{equation}
\langle q|\,\hat{\cal
K}^2\,|q\rangle\rightarrow4\langle1,1|\,\hat{\cal K}^2\,|1,1\rangle,
\end{equation}
the limiting value $\Delta{\cal K}^2_\infty$ of
$\langle1,1|\,\hat{\cal K}^2\,|1,1\rangle$ for $q\rightarrow\infty$
(without prefactors) is 2. In the considered cases $q=2$ and $q=4$,
$\Delta{\cal K}=0.72\,\Delta{\cal K}_\infty$ and $0.90\,\Delta{\cal
K}_\infty$, respectively.

For $p=-\frac{1}{2}$ - recall that ${\cal K}_{-\frac{1}{2}}$  is
proportional to $p_\ell$, see equation (31)) - and $q=2$ the matrix
elements are
\begin{eqnarray}
&&\hspace{-1cm}\langle2m+1,2n+1|\,\hat{\cal
K}_{-\frac{1}{2}}^2\,|2m+1,2n+1\rangle\propto\nonumber\\[2mm]
&&|2n+1|^{-4}|2m+1|^{-\frac{5}{4}}\left[|m|^\frac{1}{2}\left(|2m+1|^{-\frac{5}{4}}-|2m-1|^{-\frac{5}{4}}\right)\right.\nonumber\\
&&-\left.|m+1|^\frac{1}{2}\left(|2m+3|^{-\frac{5}{4}}-|2m+1|^{-\frac{5}{4}}\right)\right]\\
&&+|2m+1|^{-4}|2n+1|^{-\frac{5}{4}}\left[|n|\left(|2n+1|^{-\frac{5}{4}}-|2n-1|^{-\frac{5}{4}}\right)\right.\nonumber\\
&&-|n+1|^\frac{1}{2}\left.\left(|2n+3|^{-\frac{5}{4}}-|2n+1|^{-\frac{5}{4}}\right)\right],\nonumber
\end{eqnarray}
the sum of which is approximately 1.184, whereas for $q=4$ we have
1.689. Here the limit $q\rightarrow\infty$ is the same as for $p=0$,
namely 2. As $p_\ell$ is proportional to ${\cal K}_\frac{1}{2}$, for $q=2$ and
$q=4$ we find the ratios $\Delta p_\ell=0.77\,\Delta
p_{\ell,\infty}$ and $\Delta p_\ell=0.92\,\Delta p_{\ell,\infty}$,
respectively. Surprisingly at first sight, in the limit of vanishing
length uncertainty, the uncertainty of the conjugate momentum of
length in the form $\Delta{\cal K}_{-\frac{1}{2}}$ goes to a finite
limit, apparently violating the naive uncertainty relation, $\Delta
\ell \Delta p_\ell \geq \hbar/2$ since $\Delta \ell \Delta p_\ell
\rightarrow 0$ for large $q$.  This apparent violation may be
explained by the reformulation of the classically canonically conjugate
quantities in the quantum theory.

\section{Discussion}

The results of the present article are ``flat" solutions to the
Killing constraints, which provide a kinematic model of Minkowski
space in this (1+1)-midisuperspace, and fluctuations of
geometric quantities in these states.  We find that solutions to the Killing constraint in
its apparently most natural form are physically unacceptable -- the
expectation values of length and volume at
every atom of geometry diverge. This suggests
that the constraint, which assures flatness of space by the absence
of gravitational waves, is too strong, and that quantum theory
cannot satisfy the Killing constraints to the same extent as
classical theory: There must be fluctuations and the Killing
constraints can be valid only in some weaker form.

In Section \ref{KC} we formulate the first version of the Killing
constraint in a straight-forward way, similar to using Thiemann's
trick in the Hamiltonian constraint operator. Motivated by the
resulting divergent expectation values of volume and length we
weaken the quantum constraint with a factor of the volume.
Additional tempering is achieved by introducing the modified
operator $\hat{\left( H_K^1 \right)}_p$ of equation (\ref{Halpha})
into the construction of the constraint operator defined in equation
(\ref{Kpq}). In this way we obtain a two-parameter family $\hat{\cal
K}_{p,q}$ of volume-weighted constraints and can construct a
two-parameter family of corresponding solutions that are numerically
estimated and discussed in Section \ref{kill_c_ops}.

We find that two cases of weighting with volume are distinguished. When
$p+q=2$, the classical constraint functions are equivalent to the
original form of $\cal K$. Nevertheless, the quantum operators
$\hat{\cal K}_{p,q}$ act in different ways on SNW states. When
$p>q$, volume and length expectation values are finite and the norm
of $\hat{\cal K}|v\rangle$ is relatively small, but the
uncertainties in length and volume are divergent (i.e. $\Delta W$
diverges) . For this reason solutions of these modified constraints
are not sufficient to model quantum flat space.

With divergent uncertainties of length and volume on a single atom of geometry
we generalize to the second case with volume weighting such that $p+q>2$, a true weakening of the constraint
already at the classical level. For a growing sum $p+q$ the volume
and length uncertainties quickly decrease but
$\Delta\cal K$ grows. Notably, as $\Delta W$ becomes smaller and
smaller $\Delta\cal K$ does not grow to infinity, as one might expect from an uncertainty argument, but approaches a
finite limit, namely the value for the single fundamental SNW state
on which the solution is based ($|2,2\rangle$ in our explicit
example in Section \ref{kill_c_ops}).  The quantity $\Delta\cal K$ remains finite in the limit
$p+q\rightarrow\infty$.  The width of this residual spread of the constraint leads us to investigate
another form of the Killing constraint.

This second regularization, discussed in Section \ref{HKC}, is
motivated by the observation that, unlike the standard constraints
of canonical general relativity, the Killing constraints are not
symmetry generators of the whole theory, but physical conditions
that pick out certain states from a larger set of states. With the
physical interpretation of the classical Killing constraint $\cal K$
as the rate of change $\dot{\cal E}$ of cross section areas in the
homogenous $(x,y)$ plane times the length of an atom of geometry, it
makes sense to look for a Hermitian version of $\hat{\cal K}$. This
is done in the second version that is valid when
$X\ll\frac{4}{\mu_0}$. There is a one-parameter family of operators
${\cal K}_q$  which have essentially unique solutions  (as long as
$\mu_v$ and $\nu_v$ are integer multiples of $\mu_0$ and $\nu_0$).
These solutions are derived in Appendix \ref{HKCsol} and displayed
in equation (\ref{q_solns}).

In spite of the different regularizations, the two versions of the Killing constraint operator require tempering by
volume and lead to qualitatively similar solutions and fluctuations, although
the volume-weighted Hermitian version does not have divergent uncertainties
in geometric quantities.  Instead, length and volume on a single atom
of geometry have finite expectation values for all values of the
parameter $q>1$.  The uncertainties in geometric quantities decrease with increasing $q$.
The Hermitian Killing constraint has smaller spread in the
uncertainty of the constraint $\Delta {\cal K}$ leading to qualitative improvement.
Both solutions of the volume-weighted Killing constraints have vanishing expectation
value for the extrinsic curvatures given in Section \ref{geom_ops}.
The existence of the analytic solutions to the Hermitian Killing constraint is the most important difference
between the two versions.

We can interpret these results as follows: Metric variables like
length and volume are constructed from triad variables alone,
whereas in the present approach flatness is formulated in terms of
Killing vectors that contain connection variables. As one can expect in quantum
theory, when the constraint on one variable is relaxed, so that its
uncertainty becomes larger, the uncertainty of another variable,
which is conjugate or at least related to conjugate variables,
becomes smaller. However, the solutions to the Hermitian Killing constraint in the limit of
vanishing length (and volume) uncertainties have finite uncertainty in the
length momentum $p_\ell$. The reason is that we reformulated the
classical quantities $\ell$ and $p_\ell$ on their way from classical
expressions to well-defined operators.
In equation (\ref{length}) $\ell$ is defined in terms of holonomy
and volume operators, whereas $\hat{p}_\ell$ is defined with the
aid of the commutator $[\hat{\cal E},\hat{H}_K]$. In this way the
commutator algebra of quantum gravity operators on some
configuration spaces is not always isomorphic to the Poisson bracket
algebra of the corresponding classical quantities and quantum
uncertainty relations can deviate from a priori expectations. 
One could speculate that Planck scale modifications of
quantum uncertainties, discussed in the literature as ``Generalized
Uncertainty Principle" (see, e.g. \cite{GUPS,GUPSabine,GUP}), might
have roots in canonical quantum gravity.

Additionally, the null Killing constraints are restrictions on
time evolution so ``flatness" has a space-time character.
Hence, in carrying over perfect flatness in this sense from
classical theory to quantum theory by setting the action of the unmodified
Killing constraints on states to zero makes the uncertainty (and
even expectation values) of triad variables at every atom of
geometry infinite.

There remains much work to do to address our goal of ascertaining the effects 
of underlying fundamental geometric discreteness of LQG on the propagation of 
gravitational waves. Most immediately we need to find the physical states -- 
states that satisfy the Hamiltonian constraint -- of Minkowski space and and 
of the uni-directional wave space-times, and to show that the constraint algebra 
contains no anomalies.  Work on this is underway.

\noindent{\large\bf Acknowledgements} Work of JA was supported, in part, by the 
Gorin Foundation Fund of Hamilton College. We thank the referees for helpful
suggestions. 

\appendix

\section{Inverse volume quantization action}
\label{inv_vol}
We evaluate the action of the $\widehat{V^{-1}}$
operator on our local eigenstate $\ket{v}$. Observe that,
from equation (\ref{hA}) and the analogous identities for
the $x$ and $y$ holonomies,
\begin{equation}
\hat{h}_a^{\pm1} = \left(\frac{1}{2} \mp i\tau_a\right)\hat{e}_a^+ +
\left(\frac{1}{2} \pm i\tau_a \right)\hat{e}_a^-,
\end{equation}
where $e_a^\pm$ are operators with the following actions on our eigenstates
\begin{equation}
\hat{e}_z^\pm\left| k, \mu, \nu  \right\rangle = \left| k\pm1, \mu, \nu \right\rangle,
\end{equation}
and
\begin{equation}
\hat{e}_x^\pm\left| k, \mu, \nu \right\rangle = \left| k, \mu \pm \mu_o, \nu \right\rangle,
\end{equation}
similarly for $\hat{e}_y^{\pm}$. The following equalities hold
\begin{equation}
\left(\frac{1}{2} \pm i\tau_a\right)^2 = \left(\frac{1}{2} \pm
i\tau_a\right)
\end{equation}
and
\begin{equation}
\left(\frac{1}{2} \pm i\tau_a\right)\left(\frac{1}{2} \mp
i\tau_a\right) = 0.
\end{equation}
Also, 
\begin{equation}
\hat{V}^{\frac{1}{3}}\left| k_\pm,\mu_v, \nu_v \right\rangle =
\frac{(\gamma \ell_P^2)^{\frac{1}{2}}}{4^{\frac{1}{3}}}\sqrt[6]{|
\mu_v \nu_v k_v |}\left| k_\pm, \mu_v, \nu_v \right\rangle.
\end{equation}
For simplicity we set $\mu_0 =1=\nu_0$ in the following.
Therefore, the action of the factors are
\begin{eqnarray}
&&\hat{h}_a\left[\hat{h}_a^{-1},\hat{V}^{\frac{1}{3}}\right]\left|\bar
 a \right\rangle = \hat{h}_a\left[\left(\frac{1}{2} +
i\tau_a\right)\left(\hat{e}_a^+\hat{V}^{\frac{1}{3}}-
\hat{V}^{\frac{1}{3}}\hat{e}_a^+\right) + \left(\frac{1}{2} -
i\tau_a \right)\left(\hat{e}_a^-
\hat{V}^{\frac{1}{3}} - \hat{V}^{\frac{1}{3}}\hat{e}_a^- \right)\right]\left|\bar a \right\rangle\nonumber \\
&&= \hat{h}_a\frac{(\gamma
\ell_P^2)^{\frac{1}{2}}}{4^{\frac{1}{3}}}\sqrt[6]{|bc|}\Bigg[\left(\frac{1}{2}
+ i\tau_a\right)
\left(\sqrt[6]{|a|} - \sqrt[6]{|a+1|}\right)\left|\bar a+1 \right\rangle\nonumber  \\
&& \ \ \ \ + \left(\frac{1}{2} - i\tau_a \right)\left(\sqrt[6]{|a|} - \sqrt[6]{|a-1|}\right)\left|\bar a - 1\right\rangle \Bigg] \\
&&=\frac{(\gamma
\ell_P^2)^{\frac{1}{2}}}{4^{\frac{1}{3}}}\sqrt[6]{|bc|}\left(\left(\frac{1}{2}
-
i\tau_a\right)\hat{e}_a^+ + \left(\frac{1}{2} + i\tau_a \right)\hat{e}_a^-\right)\Bigg[\left(\frac{1}{2} + i\tau_a\right)\nonumber\\
&& \ \ \ \ \times \left(\sqrt[6]{|a|} - \sqrt[6]{|a+1|}\right)\left|
\bar a+1\right\rangle   + \left(\frac{1}{2} -
i\tau_a \right)\left(\sqrt[6]{|a|} - \sqrt[6]{|a-1|}\right)\left|\bar a -1 \right\rangle \Bigg]\nonumber \\
&&= \frac{(\gamma
\ell_P^2)^{\frac{1}{2}}}{4^{\frac{1}{3}}}\sqrt[6]{|bc|}\Bigg[\left(\frac{1}{2}
+ i\tau_a\right)\left(\sqrt[6]{|a|} - \sqrt[6]{|a+1|}\right) +
\left(\frac{1}{2} - i\tau_a \right)\left(\sqrt[6]{|a|} -
\sqrt[6]{|a-1|}\right) \Bigg] \left|\bar a\right\rangle,\nonumber
\end{eqnarray}
where $a,b,$ and $c$ stand for one of $\mu_v, \nu_v$ or $k_v$ and $\bar a$ in
the kets stands for $\mu_v$, $\nu_v$ or the pair $k_\pm$. In this
abbreviation for $|v\rangle$ the other labels are suppressed.
When $\bar a=k_\pm$, 1 is added or
subtracted from both $k_+$ and $k_-$.  In the next two
equations, where the vertex functions are eigenfunctions, we write
simply $|v\rangle$. Hence,
\begin{eqnarray}
&&\hat{h}_a\left[\hat{h}_a^{-1},\hat{V}^{\frac{1}{3}}\right]\hat{h}_b\left[\hat{h}_b^{-1},
\hat{V}^{\frac{1}{3}}\right]\hat{h}_c\left[\hat{h}_c^{-1},\hat{V}^{\frac{1}{3}}\right]\left|
v \right\rangle =
\frac{(\gamma \ell_P^2)^{\frac{3}{2}}}{4}\sqrt[3]{|k_v||\mu_v||\nu_v|} \\
&& \ \ \ \times \left[\left(\frac{1}{2} +
i\tau_3\right)\left(\sqrt[6]{|k_v|} - \sqrt[6]{|k_v+1|}\right) +
\left(\frac{1}{2} -
i\tau_3 \right)\left(\sqrt[6]{|k_v|} - \sqrt[6]{|k_v-1|}\right)\right]\nonumber \\
&& \ \ \ \times \left[\left(\frac{1}{2} +
i\tau_x\right)\left(\sqrt[6]{|\mu_v|} - \sqrt[b]{|\mu_v+1|}\right) +
\left(\frac{1}{2} - i\tau_x \right)
\left(\sqrt[6]{|\mu_v|} - \sqrt[6]{|\mu_v-1|}\right)\right]\nonumber \\
&& \ \ \  \times \left[\left(\frac{1}{2} +
i\tau_y\right)\left(\sqrt[6]{|\nu_v|} - \sqrt[6]{|\nu_v+1|}\right) +
\left(\frac{1}{2} - i\tau_y\right)\left(\sqrt[6]{|\nu_v|} -
\sqrt[6]{|\nu_v-1|}\right)\right]|v\rangle.\nonumber
\end{eqnarray}
After we take the trace, only the products including all $\tau$ terms or all
non-$\tau$ terms will remain. That is
\begin{eqnarray}
&&{\rm Tr}
\left[\hat{h}_x\left[\hat{h}_x^{-1},\hat{V}^{\frac{1}{3}}\right]
\hat{h}_y\left[\hat{h}_y^{-1},\hat{V}^{\frac{1}{3}}\right]\hat{h}_z\left[\hat{h}_z^{-1},\hat{V}^{\frac{1}{3}}\right]\right]
\left| v\right\rangle = \frac{(\gamma
\ell_P^2)^{\frac{3}{2}}}{4}\sqrt[3]{|k_v||\mu_v||\nu_v|}\nonumber \\
&& \ \ \times \Bigg(\left[\sqrt[6]{|\mu_v|} - \frac{1}{2}
\left(\sqrt[6]{|\mu_v-1|}+ \sqrt[6]{|\mu_v+1|}\right)\right]\\
&& \left[\sqrt[6]{|\nu_v|} - \frac{1}{2} \left(\sqrt[6]{|\nu_v-1|}+
\sqrt[6]{|\nu_v+1|}\right)\right] \left[\sqrt[6]{|k_v|} -
\frac{1}{2} \left(\sqrt[6]{|k_v-1|}+
\sqrt[6]{|k_v+1|}\right)\right]\nonumber \\
&&+
\frac{i}{8}\left(\sqrt[6]{|\mu_v-1|}-\sqrt[6]{|\mu_v+1|}\right)\left(\sqrt[6]{|\nu_v-1|}-\sqrt[6]{|\nu_v+1|}\right)
\left(\sqrt[6]{|k_v-1|}-\sqrt[6]{|k_v+1|}\right) \Bigg) \left| v
\right\rangle \nonumber
\end{eqnarray}
so after summing over the Levi-Civita symbol we are left with
\begin{eqnarray}&&\widehat{V^{-1}} \left| k_\pm, \mu_v,\nu_v \right\rangle =
-\frac{16\hbar^3i}{81\kappa^3\gamma^3}\frac{i(\gamma \ell_P^2)^{\frac{3}{2}}}{32}\sqrt[3]
{|k_v||\mu_v||\nu_v|}\left(\sqrt[6]{|\mu_v-1|}-\sqrt[6]{|\mu_v+1|}\right)\nonumber \\
&&\ \ \ \ \ \ \times
\left(\sqrt[6]{|\nu_v-1|}-\sqrt[6]{|\nu_v+1|}\right)\left(\sqrt[6]{|k_v -1|}-\sqrt[6]{|k_v+1|}\right) \Bigg)
\left| k_\pm, \mu_v, \nu_v  \right\rangle\nonumber \\
&&=
\frac{l_p^3\hbar^3}{162\kappa^3\gamma^{\frac{3}{2}}}\sqrt[3]{|k_v||\mu_v||\nu_v|}\left(\sqrt[6]{|\mu_v-1|}-\sqrt[6]{|\mu_v+1|}\right)
\left(\sqrt[6]{|\nu_v-1|}-\sqrt[6]{|\nu_v+1|}\right)\nonumber  \\
&&\ \ \ \ \ \ \times \left(\sqrt[6]{|k_v -1|}-\sqrt[6]{|k_v
+1|}\right) \Bigg) \left| k_\pm, \mu_v, \nu_v  \right\rangle
\end{eqnarray}
which is the result in section \ref{geom_ops} with $\mu_0 =1 = \nu_0$.

\section{Hamiltonian constraint quantization}
\label{hk1}
In $(3+1)$-LQG the kinetic part of the Hamiltonian constraint
is regularized in terms of closed loops with a ``tail."
In the first term of the Hamiltonian constraint, denoted by $H_K^1$ (as usual, $H_K$
contributes with a minus sign to the full Hamiltonian constraint)
the analogue of closed loops are composed of
point holonomies, the ``tails" are sections of adjacent edges, see
(\ref{H}) below.

Replacing the integral in the classical expression of the first part of the Hamiltonian constraint with test
function $N$,
\begin{equation}
H_K^1[N]=\frac{1}{\gamma^2}\int{\rm
d}z\,N(z)\,\frac{XE^xYE^y}{\sqrt{{\cal E}E^xE^y}},
\end{equation}
by a Riemann sum leads to terms
\begin{equation}\label{XY}\frac{1}{\gamma^2}\frac{\epsilon
X(z)Y(z)E^x(z)E^y(z)}{V(z)},
\end{equation}
where we have made use of the
volume expression $V=\sqrt{{\cal E}E^xE^y}$ and chosen $N\equiv1$.
For a representation by a Poisson bracket we take first the holonomy
along an interval to the left of a vertex, $v$
\begin{equation}
h_{z,-}=e^{\tau_3\int_-{\cal A}},
\end{equation}
where
$$\int_-=\int_{z(v)}^{z(v)-\epsilon}=-\int_{z(v)-\epsilon}^{z(v)}=:-\int_{{\cal I}_-}$$
in which $z(v)$ is the coordinate of a vertex.
To express $H_K^1$ in terms of holonomies and fluxes, we first
calculate the Poisson bracket
\begin{equation}\label{PB1}
\{h_{z,-}^{-1},V\}=\kappa\gamma\int{\rm
d}z'\,\frac{\delta}{\delta{\cal A}(z')}\,e^{+\tau_3\int_{{\cal
I}_-}{\cal A}(z){\rm d}z}\,\frac{\delta}{\delta{\cal
E}(z')}\:\epsilon\!\int{\rm d}z''V(z'').
\end{equation}
The integration in the holonomy goes in the positive direction, so
for every $z'\in{\cal I}_-$ the functional derivative with respect to
$\cal A$ gives a factor $$\tau_3\,\delta(z-z'),$$
so that the integral over $z'$ reduces to $\int_{{\cal I}_-}{\rm d}z'$.
\begin{eqnarray}\label{PB2}
\{h_{z,-}^{-1},V\}&\!\approx\!&\kappa\gamma\,\tau_3\:e^{\tau_3\int_{{\cal
I}_-}{\cal A}(z){\rm d}z}\,\int_{{\cal I}_-}{\rm
d}z'\,\frac{\delta}{\delta{\cal E}(z')}\int{\rm d}z''\sqrt{{\cal
E}(z'')E^x(z'')E^y(z'')}\nonumber\\
&\!=\!&\frac{\kappa\gamma}{2}\,\tau_3\,h_{z,-}^{-1}\:\int_{{\cal
I}_-}{\rm d}z'\,\frac{E^x(z')E^y(z')}{V(z')}.
\end{eqnarray}
Approximating the integral by the integrand at
$z$ multiplied by the interval length $\epsilon$, we finally arrive
at
\begin{equation}\label{PB5}
h_{z,-}\{h_{z,-}^{-1},V\}\approx\frac{\kappa\gamma}{2}\,\tau_3\,\frac{\epsilon\,E^x(z)E^y(z)}{V(z)}.
\end{equation}
For the interval going to the right from the vertex, the integral
$\int_+\cal A$ goes into the positive direction and
$$\frac{\delta}{\delta\cal
A}\,h_{z,+}^{-1}=-\tau_3\,h_{z,+}^{-1}.$$
In comparison with (\ref{PB5}) this gives an overall minus sign. So both Poisson
brackets give the desired classical approximation with different
signs and we can symmetrize.
\begin{eqnarray}\label{35}
\frac{\kappa\gamma}{2}\,\tau_3\,\frac{\epsilon\,E^x(z)E^y(z)}{V(z)}&\approx&\frac{1}{2}
\left(h_{z,-}\{h_{z,-}^{-1},V\}-h_{z,+}\{h_{z,+}^{-1},V\}\right)\nonumber\\
&=&-\frac{1}{2}\sum_\sigma\sigma
h_{z,\sigma}\{h_{z,\sigma}^{-1},V\},\end{eqnarray} where we have
introduced the sign factor $\sigma=\pm1$ of Ref. \cite{2}. In analogy to full LQG this
is multiplied by
\begin{equation}
\begin{array}{l}\label{36}
h_xh_yh_x^{-1}h_y^{-1}-h_yh_xh_y^{-1}h_x^{-1}=4\tau_x\sin(\mu_0X)
\sin^2\left(\displaystyle\frac{\nu_0}{2}\,Y\right)\\[3mm]
-4\tau_y\sin^2\left(\displaystyle\frac{\mu_0}{2}\,X\right)\sin(\nu_0Y)+
2\tau_3\sin(\mu_0X)\sin(\nu_0Y)
\end{array}
\end{equation}
and the trace is taken. Because of the $\tau_3$ matrix in
(\ref{35}), the only part of (\ref{36}) contributing to the trace is
$$2\tau_3\sin(\mu_0X)\sin(\nu_0Y).$$
We find
\begin{eqnarray}&&{\rm
Tr}\left[\left\{h_xh_yh_x^{-1}h_y^{-1}-h_yh_xh_y^{-1}h_x^{-1}\right\}\left(-\frac{1}{2}\right)\sum_\sigma\sigma
h_{z,\sigma}\{h_{z,\sigma}^{-1},V\}\right]\nonumber\\
&&\approx\sum_\sigma\sigma{\rm
Tr}\left[\sin(\mu_0X)\sin(\nu_0Y)\,\kappa\gamma\,\tau_3^2\,\frac{\epsilon\,E^xE^y}{V}\right]\nonumber\\
&&=-\frac{\kappa\gamma}{2}\,\frac{\epsilon\,E^x(z)E^y(z)}{V(z)}\sin(\mu_0X)\sin(\nu_0Y),
\end{eqnarray}
and in first approximation of the sine function we have
\begin{eqnarray}
&&\sum_\sigma\sigma{\rm
Tr}\left[\left\{h_xh_yh_x^{-1}h_y^{-1}-h_yh_xh_y^{-1}h_x^{-1}\right\}h_{z,\sigma}\{h_{z,\sigma}^{-1},V\}\right]\nonumber\\
&&\approx\kappa\gamma\,\mu_0\nu_0\,\frac{\epsilon\,X(z)Y(z)E^x(z)E^y(z)}{V(z)}.
\end{eqnarray}
Thus, when we replace the Poisson brackets by $(i\hbar)^{-1}$
times the commutator we obtain
\begin{equation}\label{H}\hat{H}_K^1=\frac{1}{i\hbar\kappa\gamma^3\mu_0\nu_0}\sum_\sigma\sigma{\rm
Tr}\left(\left\{\hat{h}_x\hat{h}_y\hat{h}_x^{-1}\hat{h}_y^{-1}-\hat{h}_y\hat{h}_x\hat{h}_y^{-1}\hat{h}_x^{-1}\right\}
\hat{h}_{z,\sigma}[\hat{h}_{z,\sigma}^{-1},\hat V]\right).
\end{equation}

\noindent{\bf Action on states}\\
\noindent Consider first the action of
$\hat{h}_{z,-}[\hat{h}_{z,-}^{-1},\hat{V}]$ on a vertex of a state
$T_{G,\vec k,\vec\mu,\vec\nu}$ 
of the form (\ref{T}), with the vertex function denoted shortly by
$|v\rangle$, and all edge holonomies oriented from the left to the
right.
\begin{eqnarray}\label{h-1}
\hat{h}_{z,-}^{-1}&=&\cos\left(\frac{1}{2}\int_-{\cal
A}\right)+2\tau_3\,\sin\left(\frac{1}{2}\int_-{\cal A}\right)\\
&=&\frac{1}{2}(1+2i\tau_3)\:e^{\frac{i}{2}\int_-\cal
A}+\frac{1}{2}(1-2i\tau_3)\:e^{-\frac{i}{2}\int_-\cal A}\nonumber
\end{eqnarray}
As $\int_-$ goes into the negative direction,
$e^{\frac{i}{2}\int_-\cal A}=e^{-\frac{i}{2}\int_{{\cal
I}_-}\!\!{\cal A}}$ lowers the label $k_-$ of the edge $e^-$ (to the
left of a vertex $v$) by one on the section
$(z(v)-\epsilon,z(v))\subset e^-$. For the same reason
$e^{-\frac{i}{2}\int_-\cal A}$ raises the label $k_-$ by one and so
\begin{eqnarray}
\hat{V}\hat{h}_{z,-}^{-1}|v\rangle&=&\frac{\gamma^\frac{3}{2}\ell_{\rm P}^3}{4}
\left[\frac{1}{2}(1+2i\tau_3)\sqrt{|\mu_v||\nu_v||k_v-1|}\:e^{\frac{i}{2}\int_-\cal
A}\right.\nonumber\\
&+&\left.\frac{1}{2}(1-2i\tau_3)\sqrt{|\mu_v||\nu_v||k_v+1|}\:e^{-\frac{i}{2}\int_-\cal
A}\right]|v\rangle
\end{eqnarray}
with the result that
\begin{eqnarray}\label{h-2}
\hat{h}_{z,-}[\hat{h}_{z,-}^{-1},\hat{V}]|v\rangle&=&\frac{\gamma^\frac{3}{2}\ell_{\rm P}^3}{4}\sqrt{| \mu_v \nu_v |}\times\\
&&\left[1-\frac{1}{2}\sqrt{|k_v+1|}-\frac{1}{2}\sqrt{|k_v-1|}\right.\nonumber\\
&&+\left.i\tau_3\left(\sqrt{|k_v+1|}-\sqrt{|k_v-1|}\right)\right]|v\rangle.\nonumber
\end{eqnarray}
Inserting the parts containing $\tau_3$ of (6) and (\ref{h-2}) into
(\ref{H}) gives (including $\sigma=-1$)
$$\frac{\gamma^\frac{3}{2}\ell_{\rm P}^3}{4\ell_{\rm P}^2\gamma^3\mu_0\nu_0}\sqrt{|\mu_v||\nu_v|}\left(\sqrt{|k_v+1|}-\sqrt{|k_v-1|}\right)\sin(\mu_0X)
\sin(\nu_0Y).$$ The analogous expression with $\hat{h}_{z,+}$ gives
the same as (\ref{h-2}) with the opposite sign. With both signs of
$\sigma$ we have the action on a gauge-invariant state (\ref{T})
given in equation (\ref{HK1}).

\section{Solution to the Hermitian Killing Constraint}
\label{HKCsol}

In this Appendix we work from equation (\ref{3}) to derive solutions to the Hermitian Killing constraint, equation (\ref{q_solns}).
For powers of $\hat V$ we have
\begin{equation}
\left(\hat V[{\cal
I}]\right)^q|k_\pm,\mu_v,\nu_v\rangle=\left(\frac{\gamma^\frac{3}{2}\ell_P^3}
{4}\,\sqrt{|\mu_v||\nu_v||k_v|}\right)^q|k_\pm,\mu_v,\nu_v\rangle
\end{equation}
and from equation (\ref{hx}) we may infer the action of ${\rm
Tr}\left[\tau_x{\hat h}_x^{-1}[{\cal I}]\right]$ in which $\cal I$ is understood
to contain the vertex $v$,
\begin{equation}
{\rm Tr}\left[\tau_x{\hat h}_x^{-1}[{\cal
I}]\right]|k_\pm,\mu_v,\nu_v\rangle=i\left(|k_\pm,\mu_v-\mu_0,\nu_v\rangle-|k_\pm,\mu_v+\mu_0,\nu_v\rangle\right)
\end{equation}
with an analogous equation applying for $Y$. Combining these actions
we have (writing $\mu\equiv\mu_v$ and $\nu\equiv\nu_v$)
\begin{eqnarray}\label{11}
&&\hat{\cal
K}_q|k_\pm,\mu,\nu\rangle=\frac{i(\gamma\ell_P^2)^{3q+1}}{4^{2q}}\,|k|^q\left(\frac{|\nu|^q}{\mu_0}\left[|\mu(\mu-\mu_0)|^\frac{q+1}{2}
|k_\pm,\mu-\mu_0,\nu\rangle\right.\right.\nonumber\\
&&\hspace{1cm}\left.-|\mu(\mu+\mu_0)|^\frac{q+1}{2}|k_\pm,\mu+\mu_0,\nu\rangle\right]\\
&&\hspace{1cm}\left.+\frac{|\mu|^q}{\nu_0}\left[|\nu(\nu-\nu_0)|^\frac{q+1}{2}|k_\pm,\mu,\nu-\nu_0\rangle-
|\nu(\nu+\nu_0)|^\frac{q+1}{2}|k_\pm,\mu,\nu+\nu_0\rangle\right]\right). \nonumber
\end{eqnarray}
By direct calculation it is easy to see that
\begin{equation}\label{13}
\langle k_\pm',\mu',\nu'|\,\hat{\cal
K}_q\,|k_\pm,\mu,\nu\rangle=\langle k_\pm,\mu,\nu|\,\hat{\cal
K}_q\,|k_\pm',\mu',\nu'\rangle^*;
\end{equation}
$\hat{\cal K}_q$ is a Hermitian operator.

With the superposition of vertex states of equation (\ref{8}) we can use equation (\ref{11}) to establish the following recurrence
relation involving the $a_{\mu,\nu}$ coefficients
\begin{eqnarray}\label{21}
0&=&|k_v|^q\left(\frac{|\nu|^q}{\mu_0}\left[|\mu(\mu+\mu_0)|^\frac{q+1}{2}a_{\mu+\mu_0,\nu}-
|\mu(\mu-\mu_0)|^\frac{q+1}{2}a_{\mu-\mu_0,\nu}\right]\right.\nonumber\\
&&\left.+\frac{|\mu|^q}{\nu_0}\left[|\nu(\nu+\nu_0)|^\frac{q+1}{2}a_{\mu,\nu+\nu_0}-|\nu(\nu-\nu_0)|^\frac{q+1}{2}a_{\mu,\nu-\nu_0}\right]\right).
\end{eqnarray}
Obviously, there exists a trivial solution, where $k_v\equiv
k_++k_-=0$. If we seek nontrivial solutions, we may safely assume
$k_+\neq-k_-$ and divide by $|k_v|^q$. Similarly, observe that in
the case that either $\mu=0$ or $\nu=0$, the recurrence relation
given in equation (\ref{21}) is again trivially satisfied
\footnote{Provided $q\neq1$. However if this is the case, then
setting $\mu=0$ and $\nu\neq0$ (for instance) reduces eq. (\ref{21})
to $a_{\mu+\mu_0,\nu}=a_{\mu-\mu_0,\nu}$, which clearly admits no
normalizable, nondegenerate solutions.}, meaning if we seek
additional solutions we may divide by $|\mu\nu|^q$
\begin{eqnarray}\label{9}
0&=&\frac{1}{\mu_0}\left(|\mu|^\frac{1-q}{2}|\mu+\mu_0|^\frac{1+q}{2}a_{\mu+\mu_0,\nu}-
|\mu|^\frac{1-q}{2}|\mu-\mu_0|^\frac{1+q}{2}a_{\mu-\mu_0,\nu}\right)\nonumber\\[2mm]
&+&\frac{1}{\nu_0}\left(|\nu|^\frac{1-q}{2}|\nu+\nu_0|^\frac{1+q}{2}a_{\mu,\nu+\nu_0}-
|\nu|^\frac{1-q}{2}|\nu-\nu_0|^\frac{1+q}{2}a_{\mu,\nu-\nu_0}\right).
\end{eqnarray}
The standard separation ansatz $a_{\mu,\nu}=\alpha_\mu\beta_\nu$
leads to
\begin{equation}
\label{alpha_eqn}
-\mu_0C=|\mu|^\frac{1-q}{2}\left(|\mu+\mu_0|^\frac{1+q}{2}\alpha_{\mu+\mu_0}-|\mu-\mu_0|^\frac{1+q}{2}\alpha_{\mu-\mu_0}\right)
\end{equation}
and
\begin{equation}
\label{beta_eqn}
\nu_0C=|\nu|^\frac{1-q}{2}\left(|\nu+\nu_0|^\frac{1+q}{2}\beta_{\nu+\nu_0}-|\nu-\nu_0|^\frac{1+q}{2}\beta_{\nu-\nu_0}\right).
\end{equation}
From the first of these relations we obtain
\begin{equation}
\alpha_{\mu+\mu_0}=\left|\frac{\mu-\mu_0}{\mu+\mu_0}\right|^\frac{q+1}{2}\!\!\alpha_{\mu-\mu_0}-\frac{C\,\mu_0
|\mu|^\frac{q-1}{2}}{|\mu+\mu_0|^\frac{q+1}{2}}.
\end{equation}
Iteration yields
\begin{equation}\label{5}
\alpha_{\mu+(2m+1)\mu_0}=\left|\frac{\mu-\mu_0}{\mu+(2m+1)\mu_0}\right|^\frac{q+1}{2}\!\!\!\!\alpha_{\mu-\mu_0}-
\frac{\mu_0C}{|\mu+(2m+1)\mu_0|^\frac{q+1}{2}}\sum_{l=0}^m|\mu+2l\mu_0|^\frac{q-1}{2}.
\end{equation}
For large $m$  the finite sum can be approximated by
$$(2\mu_0)^\frac{q-1}{2}\sum_{l=0}^ml^\frac{q-1}{2},$$
the leading term of which for $q\geq1$ can be approximated by an
integral
$$(2\mu_0)^\frac{q-1}{2}\,\frac{2m^\frac{q+1}{2}}{q+1}.$$
So the second term in (\ref{5}) is, in leading order, equal to
$$-\frac{C}{q+1},$$
i.e. for large $m$ it goes to a constant
as does $\alpha_{\mu+(2m+1)\mu_0}$. As a consequence, for $C\neq0$
we do not find nontrivial normalizable states, and we have to
consider only the case $C=0$. With the reparametrizations
$\mu\rightarrow\bar\mu+\mu_0$, where $\bar\mu$ is some initial
value, and $m\rightarrow m-1$ we obtain from the first part of
equation (\ref{5})
\begin{equation}\label{6}
\alpha_{\bar\mu+2m\mu_0}=\left|\frac{\bar\mu}{\bar\mu+2m\mu_0}\right|^\frac{q+1}{2}\!\!\!\alpha_{\bar\mu}.
\end{equation}
For some given initial value $\alpha_{\bar\mu}\neq0$ the series
$$\sum_{m=0}^\infty|\alpha_{\bar\mu+2m\mu_0}|^2$$
converges, when $q>0$.

For a solution of the constraint equation we must also consider
decreasing indices of $\alpha$. Analogous to equation (\ref{6}),
\begin{equation}\label{7}
\alpha_{\bar\mu-2m\mu_0}=\left|\frac{\bar\mu}{\bar\mu-2m\mu_0}\right|^\frac{q+1}{2}\!\!\!\alpha_{\bar\mu}.
\end{equation}

The simplest nontrivial solution is obtained by choosing an initial
value $\bar\mu=\mu_0$. Then from $m=1$ we find
\begin{equation}
\alpha_{-\mu_0}=\alpha_{\mu_0},
\end{equation}
and
\begin{equation}
\alpha_{\pm(2m+1)\mu_0}=\left|\frac{1}{2m+1}\right|^\frac{q+1}{2},
\end{equation}
a symmetric series in $m$.

For even multiples of $\mu_0$ we find from equation (\ref{6}) that
$\alpha_{2\mu_0}=0$ (and $\alpha_{2m\mu_0}=0$ for all
$m$, irrespectively of the value of $\alpha_0$). On the other hand,
if $\alpha_{-2\mu_0}\neq0$, $\alpha_0$ diverges. As a consequence,
all states with even multiples of $\mu_0$ are excluded and we can
also set $\alpha_0=0$, thus avoiding zero vacuum and zero length
states, in accordance with the division by $|\mu\nu|^q$ in
equation (\ref{9}).

Under the assumption that $\bar\mu$ is an integer multiple of
$\mu_0$, only the odd multiples are nonzero. In the following we
make use of the abbreviation
\begin{equation}
\alpha_{2m+1}\equiv\alpha_{(2m+1)\mu_0}.
\end{equation}

For finite volume and length expectation values, and for finite
second moments, both
$$\sum_m\sqrt{|2m+1|}\,|\alpha_{2m+1}|^2$$
and
$$\sum_m|2m+1|\,|\alpha_{2m+1}|^2$$
must converge. From the first condition follows
\begin{equation}
q>\frac{1}{2},
\end{equation}
and from the second
\begin{equation}
q>1.
\end{equation}
So as with the commutator version, the Hermitian Killing operator
$\hat{\cal K}$ must be volume weighted to give rise to finite length and
volume expectation values and uncertainties.

To finish the solution we observe that the coefficients $\beta_n$
are found in the same way and starting with an initial value
$\beta_1=1$ we have
\begin{equation}
\beta_{\pm(2n+1)}=\left|\frac{1}{2n+1}\right|^\frac{q+1}{2}.
\end{equation}
Combining these results we find coefficients
$a_{2m+1,2n+1}=\alpha_{2m+1}\,\beta_{2n+1}$ for a solution with
initial coefficient $a_{1,1}$
\begin{equation}\label{amn}
a_{2m+1,2n+1}=\frac{a_{1,1}}{|(2m+1)(2n+1)|^\frac{q+1}{2}}.
\end{equation}
The norm square  of this solution state to the Hermitian Killing constraint of equation (\ref{8}), denoted by
$|q\rangle$, is
\begin{equation}
\langle
q|q\rangle=\sum_{m,n=-\infty}^\infty|a_{2m+1,2n+1}|^2=4a_{1,1}^2\sum_{m=0}^\infty\frac{1}{(2m+1)^{q+1}}\sum_{n=0}^\infty\frac{1}{(2n+1)^{q+1}},
\end{equation}
so that
\begin{equation}
|\!|\;|q\rangle\,|\!|=2a_{1,1}\sum_{n=0}^\infty\frac{1}{(2n+1)^{q+1}}.
\end{equation}
This sum can be written as
$$\sum_{n=0}^\infty\frac{1}{(2n+1)^{q+1}}+\sum_{n=1}^\infty\frac{1}{(2n)^{q+1}}-\sum_{n=1}^\infty\frac{1}{(2n)^{q+1}}=\sum_{n=1}^\infty
\frac{1}{n^{q+1}}-\frac{1}{2^{q+1}}\sum_{n=1}^\infty\frac{1}{n^{q+1}}$$
and expressed in terms of the Riemann zeta function. Thus,
\begin{equation}
|\!|\;|q\rangle\,|\!|=a_{1,1}\,\frac{2^{q+1}-1}{2^q}\,\zeta(q+1).
\end{equation}
For
\begin{equation}
a_{1,1}=\frac{2^q}{2^{q+1}-1}\,\frac{1}{\zeta(q+1)}
\end{equation}
we have a normalized state with finite expectation values
\begin{equation}
|q\rangle=\frac{2^q}{2^{q+1}-1}\frac{1}{\zeta(q+1)}\sum_{m,n=-\infty}^\infty\frac{1}{|(2m+1)(2n+1)|^\frac{q+1}{2}}\,
|2m+1,2n+1\rangle,
\end{equation}
where we have used the notation of equation (\ref{statenotation}). In the limit of large $q$ the state $|q\rangle$ goes to
a superposition of four states with the same length and volume
expectation values,
\begin{equation}
\lim_{q\rightarrow\infty}|q\rangle\longrightarrow\frac{1}{2}\left(|1,1\rangle+|1,-1\rangle+|-1,1\rangle+|-1,-1\rangle\right).
\end{equation}

Next we calculate the expectation value and the uncertainty of
$\sqrt{|\mu\nu|}$ (up to a factor $\sqrt{\mu_0\nu_0}$), which is
contained in both length and volume,
\begin{equation}
W:=\langle\sqrt{|\mu\nu|}\rangle/\sqrt{\mu_0\nu_0}=4\sum_{m,n=0}^\infty
a_{2m+1,2n+1}^2\sqrt{(2m+1)(2n+1)},
\end{equation}
the factor 4 coming from the four quadrants in the $(m,n)$ plane.
With the coefficients inserted from equation (\ref{amn})
\begin{equation}
W=2\,\left(\frac{2^{q+\frac{1}{2}}
-1}{2^{q+1}-1}\right)^2\left(\frac{\zeta(q+\frac{1}{2})}{\zeta(q+1)}\right)^2.
\end{equation}
In the same way we calculate
\begin{equation}
\langle |\mu\nu |\rangle=4\,\mu_0\nu_0\left(\frac{2^q-1}{2^{q+1}-1}\right)^2\left(\frac{\zeta(q)}{\zeta(q+1)}\right)^2
\end{equation}
and
\begin{equation}
\Delta
W=\left(\langle |\mu\nu | \rangle-\langle\sqrt{|\mu\nu|}\rangle^2\right)^\frac{1}{2}/\sqrt{\mu_0\nu_0}
\end{equation}
which gives equation (\ref{HKCDW}),
\begin{equation}
\Delta
W=2\,\frac{[(2^q-1)^2(2^{q+1}-1)^2\zeta(q)^2\zeta(q+1)^2-(2^{q+\frac{1}{2}}-1)^4\zeta(q+\frac{1}{2})^4]^\frac{1}{2}}
{(2^{q+1}-1)^2\,\zeta(q+1)^2}.
\end{equation}
In accordance with equation (\ref{29}) the length and volume uncertainties go
to zero for large $q$.


\begin{thebibliography}{99}

\bibitem{RSareavol} C. Rovelli and L. Smolin {\em Nucl. Phys.} (1995)
{\bf B 442} 593; erratum {\em Nucl. Phys.} {\bf B 456} (1995) 753.

\bibitem{Lvol} R. Loll
``Spectrum of the Volume Operator in Quantum Gravity"
{\em Nucl. Phys.} {\bf B460} (1996) 143-154
arXiv:gr-qc/9511030.

\bibitem{ALareavol} A. Ashtekar and J. Lewandowski
``Quantum Theory of Gravity I: Area Operators"
{\em Class. Quant. Grav.} {\bf 14} (1997) A55-A82
arXiv:gr-qc/9602046;
A. Ashtekar and J. Lewandowski
``Quantum theory of geometry. II: Volume operators"
{\em Adv. Theor. Math. Phys.} {\bf 1} (1998) 388
arXiv:gr-qc/9711031.

\bibitem{Tlength} T. Thiemann
``A length operator for canonical quantum gravity"
 {\em J. Math. Phys.} {\bf 39} (1998) 3372-3392
 arXiv:gr-qc/9606092.

 \bibitem{Mangle} S. Major,
``Operators for quantized directions''
{\em Class.  Quant.  Grav.} {\bf 16} (1999) 3859
arXiv:gr-qc/9905019.

\bibitem{Blength} E. Bianchi
``The length operator in Loop Quantum Gravity"
{\em  Nucl. Phys.} {\bf B 807} (2009) 591-624, arXiv:0806.4710.

\bibitem{HM} F. Hinterleitner and S. Major, ``On plane gravitational
waves in real connection variables", {\em Phys. Rev.} {\bf D 83}
044034, arXiv:1006.4146.

\bibitem{HM2} F. Hinterleitner and S. Major, ``Toward Loop Quantization of Plane Gravitational Waves"
{\em Class.~Quantum Grav.} {\bf 29} (2012) 065019, arXiv:1106.1448.

\bibitem{BD} K. Banerjee and G. Date ``Loop Quantization of Polarized Gowdy Model on $T^3$:
Classical Theory", {\em Class. Quantum Grav.} {\bf25} (2008) 105014, arXiv:0712.0683,\\ K. Banerjee and G. Date, 
``Loop Quantization of Polarized Gowdy Model on $T^3$: Kinematical States and Constraint
Operators"
{\em Class. Quantum Grav.} {\bf 25} (2008) 145004,
arXiv:0712.0687.

\bibitem{2} M. Bojowald and R. Swiderski, ``Spherically Symmetric Quantum Geometry:
Hamiltonian Constraint," {\em Class. Quantum Grav.} {\bf 23} (2006) 2129-2154, arXiv:gr-qc/ 0511108.

\bibitem{N} D. Neville, {\em Class. Quantum Grav.} {\bf 10} (1993) 2223; {\em Phys. Rev.} {\bf D 55} (1997) 766; {\em Phys. Rev.} {\bf D 55} (1997) 2069; {\em Phys. Rev.} {\bf D 56} (1997) 3485; {\em Phys. Rev.} {\bf D 57} (1998) 986.

\bibitem{Bweaves} R. Borissov, {\em Phys. Rev.} {\bf D 49} (1994) 923.

\bibitem{B} C. Beetle,
``Midi-Superspace Quantization of Non-Compact Toroidally Symmetric Gravity"
{\em Adv. Theor. Math. Phys.} {\bf 2} (1998) 471-495
arXiv:gr-qc/9801107.

\bibitem{MM} G. A. Mena Marugan and M. Montejo,
``Quantization of pure gravitational plane waves"
{\em Phys. Rev.} {\bf D 58} (1998) 104017
arXiv:gr-qc/9806105.

\bibitem{5} T. Thiemann, ``Quantum Spin Dynamics (QSD)", {\em Class.
Quantum Grav.} {\bf15} (1998) 139-173, arXiv:gr-qc/960608.

\bibitem{OT} T. Thiemann and O. Winkler, ``Gauge Field Theory Coherent States 
(GCS) : IV. Infinite Tensor Product and Thermodynamical Limit", {\em Class. 
Quantum Grav.} {\bf 18} (2001) 4997-5054, arXiv:hep-th/0005235;

\bibitem{grif} J. B. Griffiths and J. Podolsk\'{y} 2009 {\em Exact
Space-Times in General Relativity} (Cambridge: Cambridge University
Press).

\bibitem{cos} M. Bojowald, ``Isotropic Loop Qauntum Cosmology", {\em
Class. Quantum Grav.} {\bf19} (2002) 2717-2742, arXiv:gr-qc/0202077,\\ M.
Bojowald, ``Loop Quantum Cosmology", Living Rev. Relativity, 11
(2008) 4, http://www.livingreviews.org/lrr-2008-4.

\bibitem{8} T. Thiemann, {\em Modern Canonical Quantum General Relativity} (Cambridge, 2007).

\bibitem{MTW} C. Misner, K. Thorne, J. Wheeler,
{\em Gravitation}
(W. H. Freeman and Company, New York, 1973), section 35.9.

\bibitem{GUPS} F. Scardigli, ``Generalized Uncertainty Principle in Quantum Gravity from Micro-Black Hole Gedanken Experiment"
{\em Phys. Lett.} {\bf B452} (1999) 39, arXiv:hep-th/9904025.

\bibitem{GUPSabine} S. Hossenfelder, ``Minimal Length Scale Scenarios for Quantum Gravity"
{\em Living Rev. Relativity} {\bf 16} (2013) 2, arXiv:1203.6191.

\bibitem{GUP} B. Carr, L. Modesto and I. Pr\'{e}mont-Schwarz,
``Generalized Uncertainty Principle and Self-dual Black Holes",
arXiv:1107.0708.
\end{thebibliography}
\end{document}